\documentclass[pdflatex,sn-mathphys-num]{sn-jnl}
\usepackage[utf8]{inputenc}
\usepackage{bm}
\usepackage{amsfonts}
\usepackage{amssymb}
\usepackage{amsmath}
\usepackage{algorithm}
\usepackage{algpseudocode}
\usepackage{graphicx}
\usepackage{siunitx}
\usepackage{multicol}

\begin{document}

% Copyright
\onecolumn
\thispagestyle{empty}
\Huge Copyright Notice \\

\large
© The Author(s) 2025. \textbf{Open Access} This article is licensed under a Creative Commons Attribution-NonCommercial-NoDerivatives 4.0 International License, which permits any non-commercial use, sharing, distribution and reproduction in any medium or format, as long as you give appropriate credit to the original author(s) and the source, provide a link to the Creative Commons licence, and indicate if you modified the licensed material. You do not have permission under this licence to share adapted material derived from this article or parts of it. The images or other third party material in this article are included in the article’s Creative Commons licence, unless indicated otherwise in a credit line to the material. If material is not included in the article’s Creative Commons licence and your intended use is not permitted by statutory regulation or exceeds the permitted use, you will need to obtain permission directly from the copyright holder. To view a copy of this licence, visit \href{http://creativecommons.org/licenses/by-nc-nd/4.0/}{http://creativecommons.org/licenses/by-nc-nd/4.0/}.

\vfill
This work has been accepted for publication in \textit{EURASIP Journal on Wireless Communications and Networking} (Volume 2025, Article Number 17, March 2025). The final published version is available via Springer Nature Link, DOI: \href{https://doi.org/10.1186/s13638-025-02445-0}{10.1186/s13638-025-02445-0}.

\clearpage
\normalsize

% Article
\title{Robust Design of Reconfigurable Intelligent Surfaces for Parameter Estimation in MTC}

\author*[1,2]{\fnm{Sergi} \sur{Liesegang}}\email{sergi.liesegang@unicas.it}

\author[3]{\fnm{Antonio} \sur{Pascual-Iserte}}\email{antonio.pascual@upc.edu}

\author[3]{\fnm{Olga} \sur{Mu\~noz}}\email{olga.munoz@upc.edu}

\affil[1]{\orgdiv{Department of Electrical and Information Engineering}, \orgname{University of Cassino and Southern Lazio}, \orgaddress{\postcode{03043}, \city{Cassino}, \country{Italy}}}

\affil[2]{\orgname{Consorzio Nazionale Interuniversitario per le Telecomunicazioni}, \orgaddress{\postcode{43124}, \city{Parma}, \country{Italy}}}

\affil[3]{\orgdiv{Department of Signal Theory and Communications}, \orgname{Universitat Polit\`ecnica de Catalunya}, \orgaddress{\postcode{08034}, \city{Barcelona}, \country{Spain}}}

\abstract{This paper introduces a reconfigurable intelligent surface (RIS) to support parameter estimation in machine-type communications (MTC). We focus on a network where single-antenna sensors transmit spatially correlated measurements to a multiple-antenna collector node (CN) via non-orthogonal multiple access. We propose an estimation scheme based on the minimum mean square error (MMSE) criterion. We also integrate successive interference cancelation (SIC) at the receiver to mitigate communication failures in noisy and interference-prone channels under the finite blocklength (FBL) regime. Moreover, recognizing the importance of channel state information (CSI), we explore various methodologies for its acquisition at the CN. We statistically design the RIS configuration and SIC decoding order to minimize estimation error while accounting for channel temporal variations and short packet lengths. To mirror practical systems, we incorporate the detrimental effects of FBL communication and imperfect CSI errors in our analysis. Simulations demonstrate that larger reflecting surfaces lead to smaller MSEs and underscore the importance of selecting an appropriate decoding order for accuracy and ultimate performance.}


\maketitle

\section{Introduction} \label{sec:1}
Machine-type communications (MTC) have become pivotal for the advancement of mobile generations \cite{Che17}. They represent systems where groups of simple devices non-orthogonally transmit information to a base station (BS) or collector node (CN) with minimal to no human oversight \cite{Boc18}. Applications of MTC include health monitoring, location tracking, and smart metering, among others. In scenarios involving sensors measuring specific parameters (e.g., temperature), the CN estimates sensed data based on potentially noisy observations from these devices. Due to the spatial density of terminals, data correlation is significant \cite{Sha13}, allowing for improved accuracy through the use of appropriate estimators \cite{Kay93}.

In instances where the channel quality between sensors and the CN is poor, transmitting measurements from MTC devices can pose challenges. Examples include setups with (i) strong Rayleigh or Rician fading with a weak line of sight (LoS) \cite{Tse05}, as well as (ii) significant propagation losses in millimeter wave (mmWave) \cite{Hem18} or terahertz (THz) \cite{Pol20} bands. Such conditions result in excessively high decoding error probabilities, rendering reliable communication uncertain. Moreover, the need for retransmissions in case of transmission failure could exacerbate complexity, latency, and power consumption, which are the most critical factors in the majority of MTC networks \cite{Boc18}. Finally, given that MTC data packets are typically short, classical Shannon metrics overestimate performance, and finite blocklength (FBL) analysis must be used instead \cite{Pol10}. {\color{black} This is especially true in applications such as narrowband Internet of Things (NB-IoT) \cite{3GPP45820}, where the number of time-frequency resources is quite limited. Additionally, the lack of decoding error of the infinite packets assumption is unrealistic and must be disregarded (there is always a non-zero probability of communication failure).}

This paper delves into leveraging reconfigurable intelligent surfaces (RIS) to enhance system performance \cite{DiR20, Wan21}. RISs are expansive passive surface structures capable of adapting to the wireless environment. Functioning as reflectors, they can redirect signals towards target destinations to amplify signal strength. With attributes such as received power gain, high scalability, low cost, and ease of deployment, RISs emerge as promising technologies for future cellular systems \cite{Raj21}.

In this context, channel state information (CSI) becomes indispensable for achieving substantial beamforming gains. However, due to the passive nature of RISs and their numerous elements, channel estimation poses a formidable challenge \cite{Wei21}. Consequently, we will explore various strategies for acquiring this crucial knowledge feasibly. Accordingly, the RIS will be configured to minimize parameter estimation errors while explicitly considering the impact of FBL communication and imperfect CSI (I-CSI) errors. The design will heavily rely on statistical information, ensuring robustness against the aforementioned uncertainties over the long term.

{\color{black} Given the extensive connectivity in MTC networks, the available resources are insufficient for orthogonal transmission. In other words, the scarcity of electromagnetic spectrum forces devices to share all resources. This situation aggravates even more when the number of sensors increases, i.e., massive MTC (mMTC). Consequently, to keep the analysis general (schemes with such reuse and interference), resorting to non-orthogonal multiple access (NOMA) becomes imperative \cite{Liu17, Yan19, Elb20}.} 

As a result, the signals received from different sensors are also susceptible to interference. We will explore successive interference cancelation (SIC) as a decoding procedure to address this issue and examine various proposals for selecting the decoding order \cite{Ver98, Dai18, Din20}. In this context, the RIS plays a crucial role in adapting the channel to the SIC procedure, thereby significantly mitigating the impact of interference and alleviating channel quality drawbacks \cite{Li21, Liu22}.

\subsection{Prior work} \label{sec:1.1}
During the past few years, RISs have garnered extensive attention from both academic and industrial communities, with recent studies demonstrating their potential to substantially enhance wireless network performance \cite{DiR20, Bjor20}. 

However, most existing works primarily focus on data rate optimization. In \cite{Cao21}, the authors optimized the sum rate with respect to (w.r.t.) the RIS response in device-to-device communications. They derived BS beamforming, power allocation, and user pairing by using block coordinate descent methods. Similarly, \cite{Mu20} explored the use of RIS to support NOMA-based transmissions with SIC decoding schemes, employing convex approximations and relaxations to jointly design RIS and BS configurations that maximize the total data rate. \cite{Xu21} characterized the ergodic rates in the presence of correlated Rician fading and subsequently optimized them using \textit{alternating optimization} (AO) techniques. 

{\color{black} AO is also exploited in \cite{Pen22}, which addresses the joint optimization of the BS transmit beamforming and the RIS phase shifts in a 6G system with hardware impairments and I-CSI. By leveraging Bernstein-type inequality, the authors minimize the BS transmit power subject to outage probability and unit-modulus constraints. Another robust approach is tackled in \cite{Yao23}, where the authors design the precoders and phase-shifters to maximize the worst-case sum rate in a RIS-aided cell-free architecture under CSI uncertainties and backhaul limitations. Given the complexity of the resulting formulation, methods such as the penalty convex-concave procedure and successive convex approximation are ultimately applied to find a feasible solution. Likewise, the authors in \cite{Che23} delve into multi-RIS-assisted networks and propose a robust and flexible scheme balancing beamforming gain and I-CSI error. Based on that, the BS precoding, channel estimation error, and RIS deployment are jointly optimized to maximize the weighted sum rate. The inherent nonconvex nature of the problem is circumvented via fractional programming plus a gradient-projected AO algorithm.

Last, the authors of \cite{Rab24} concentrated on the impact of non-linearities when modeling the wireless channel in RIS-aided environments. To accurately the input-output relationship, a novel mathematical framework was proposed. Such formulation evinced the performance gap between lineal and non-lineal assumptions in real-world scenarios.} To the best of the authors' knowledge, at the time of writing, no other studies have investigated the utilization of RIS for parameter estimation in MTC systems.

On the other hand, the literature discusses various approaches to acquiring knowledge on RIS channels. For example, in \cite{Yan20}, the authors proposed an estimation and transmission protocol based on orthogonal frequency-division multiplexing (OFDM) and overhead reduction for RIS-assisted scenarios with frequency-selective channels. 

\noindent
The issue of I-CSI was also addressed in \cite{Hua20}, where deep reinforcement learning was employed for channel estimation. Similarly, in \cite{Zhi23}, the authors investigated the use of linear estimators to obtain the instantaneous channel (utilized for BS spatial filters), while the RIS matrix was designed based on channel distribution information (CDI).

\subsection{Contributions} \label{sec:1.2}
\begin{figure}[t]
    \centerline{\includegraphics[scale = 1]{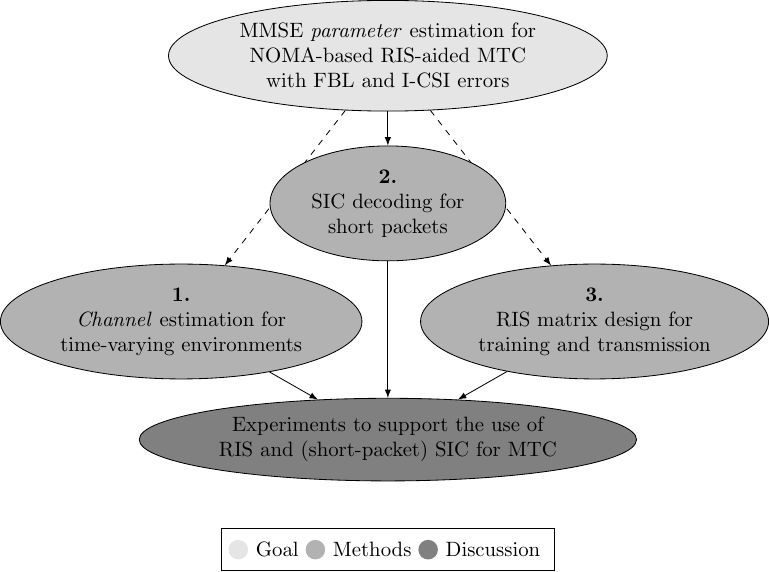}}
	\caption{\color{black} Summary of main contributions.}
	\label{fig:0}
 \end{figure}

{\color{black} As illustrated in Fig.~\ref{fig:0}, the main contributions are listed below:}
\begin{itemize}
    \setlength\itemsep{0em}
    \item A minimum mean square error (MMSE) estimation scheme for NOMA-based MTC in RIS-aided scenarios, with FBL communication and I-CSI errors.
    \item Feasible estimation protocols for acquiring the channel knowledge in time-varying environments.
    \item A tractable decoding strategy based on (short-packet) SIC and criteria for selecting the decoding order.
    \item An iterative algorithm for designing the reflection matrix during both training and transmission phases.
    \item Numerical results that justify employing RIS and (short-packet) SIC to enhance MTC transmissions.
\end{itemize}

\subsection{Organization} \label{sec:1.3}
The remainder of this paper is structured as follows. The description of the methods is split throughout Sections~\ref{sec:2} to \ref{sec:5}. Section~\ref{sec:2} characterizes the system model. Section~\ref{sec:3} defines the channel estimation framework and the resulting SIC decoding procedure under the FBL regime is studied in Section~\ref{sec:4}. Section~\ref{sec:5} is devoted to the robust design of the RIS. The results are presented in Section~\ref{sec:6} along with the discussion of the simulation experiments. Final thoughts and concluding remarks are given in Section~\ref{sec:7}.

\subsection{Notation} \label{sec:1.4}
In this work, scalars are denoted by italic letters. Boldface lowercase and uppercase letters represent vectors and matrices, respectively. The zero and all-ones vectors of length $m$ are denoted by $\mathbf{0}_m$ and $\mathbf{1}_m$, respectively. $\mathbf{I}_m$ denotes the identity matrix of size $m \times m$. $\mathbb{R}^{m \times n}$ and $\mathbb{C}^{m \times n}$ denote the $m$ by $n$ dimensional real and complex spaces, respectively. The Hadamard (or element-wise) and Kronecker products are denoted by $\odot$ and $\otimes$, respectively. The transpose, Hermitian, inverse, and trace operators are denoted by $(\cdot)^{\textrm{T}}$, $(\cdot)^{\textrm{H}}$, $(\cdot)^{\textrm{-1}}$, and $\textrm{tr}(\cdot)$, respectively. The expectation operator is denoted by $\mathbb{E}[\cdot]$. The real and complex proper Gaussian distributions are denoted by $\mathcal{N}(\cdot,\cdot)$ and $\mathcal{CN}(\cdot,\cdot)$, respectively.

\section{Methods: System model} \label{sec:2}
We consider a scenario similar to that in \cite{Lie23}, where $M$ single-antenna sensors are connected to a CN equipped with $K$ antennas. Each sensor observes a parameter $\theta_i \in \mathbb{R}$ subject to measurement noise $\eta_i \in \mathbb{R}$ with $i \in \{1,\ldots,M\}$. Thus, the information to be transmitted is given by \cite[(1)]{Lie21}:
\begin{equation}
    \mathbf{x} = \bm{\theta} + \bm{\eta} \in \mathbb{R}^M,
    \label{eq:1}
\end{equation}
where $\mathbf{x} = [x_1,\ldots,x_M]^{\textrm{T}}$ contains the different observations, $\bm{\theta} = [\theta_1,\ldots,\theta_M]^{\textrm{T}} \in \mathbb{R}^M$ is the parameter vector, Gaussian distributed with mean $\bm{\mu}_{\bm{\theta}} \in \mathbb{R}^M$ and covariance matrix $\mathbf{C}_{\bm{\theta}} \in \mathbb{R}^{M \times M}$, i.e., $\bm{\theta} \sim \mathcal{N}(\bm{\mu}_{\bm{\theta}},\mathbf{C}_{\bm{\theta}})$, and $\bm{\eta} = [\eta_1,\ldots,\eta_M]^{\textrm{T}} \in \mathbb{R}^M$ is the set of observation noises, mutually independent and Gaussian distributed with zero mean and covariance matrix $\mathbf{C}_{\bm{\eta}} \in  \mathbb{R}^{M \times M}$, i.e., $\bm{\eta} \sim \mathcal{N}(\mathbf{0}_M,\mathbf{C}_{\bm{\eta}})$. Denoting the noise powers as $\sigma_{\eta_i}^2$, we have $\mathbf{C}_{\bm{\eta}} = \textrm{diag}(\sigma_{\eta_1}^2,\ldots,\sigma_{\eta_M}^2)$. 

Accordingly, each sensor maps the individual $x_i \in \mathbb{R}$ into a transmit symbol $s_i \in \mathbb{C}$ through an encoder $\mathcal{G}_i(\cdot)$, i.e., $s_i \triangleq \mathcal{G}_i(x_i)$. We further assume these symbols have zero means and transmit powers $P_i$. Note that the function $\mathcal{G}_i(\cdot)$ includes the error correcting code as well as the modulation scheme, namely, the modulation and coding scheme (MCS). {\color{black}
In that sense, even though the transmit symbols $s_i$ originate from the parameter observations $x_i$ in \eqref{eq:1}, which are indeed correlated, the coding scheme can transform the statistics to achieve independent values before actual communication.}

Considering shared resources (i.e., NOMA transmission), the received signal at the CN yields:
\begin{equation}
    \mathbf{y} \triangleq \sum_{i = 1}^M \mathbf{q}_i s_i + \mathbf{w} \in \mathbb{C}^K,
    \label{eq:2}
\end{equation}
where $\mathbf{q}_i \in \mathbb{C}^K$ is the direct channel between sensor $i$ and the CN, and $\mathbf{w} \sim \mathcal{CN}(\mathbf{0}_K,\sigma_w^2\mathbf{I}_K)$ is the complex additive white Gaussian noise (AWGN).

Under the use of a linear filter $\mathbf{f}_i \in \mathbb{C}^K$ for detecting sensor $i$, the signal-to-interference-plus-noise ratio (SINR) of its received signal is given by \cite[(2)]{Cao21}:
\begin{equation}
    \rho_i = \frac{P_i \vert \mathbf{f}_i^{\textrm{H}} \mathbf{q}_i \vert^2 }{\displaystyle \sigma_w^2 \| \mathbf{f}_i \|_2^2 + \sum_{j \neq i} P_j \vert \mathbf{f}_i^{\textrm{H}} \mathbf{q}_j \vert^2}.
    \label{eq:3}
\end{equation}

As previously mentioned, despite the beamforming design discussed in Section V, in real scenarios, the received power $P_i \vert \mathbf{f}_i^{\textrm{H}} \mathbf{q}_i \vert^2$ can be relatively small compared to the noise and interference levels. Therefore, in this paper, we incorporate an RIS to support the transmission from the sensors to the CN. Using phase shifters, the RIS will direct the different signals toward the CN as needed \cite{Hua19, Bjo20}. 

With the above considerations, the response of an RIS with $L$ elements is given by\footnote{\color{black} When non-linearities arise due to mutual coupling, this linear model for the RIS is no longer valid \cite{Rab24}. However, as discussed later on, our design can be extended to include such effects. For simplicity, we stick to this case and leave the corresponding analysis for future work.} \cite{He20, Bjo21}:
\begin{equation}
    \bm{\Psi} \triangleq \textrm{diag}\left(\lambda_1 e^{j\phi_1},\ldots,\lambda_L e^{j\phi_L}\right) \in \mathbb{C}^{L \times L},
    \label{eq:4}
\end{equation}
where $\lambda_l \in \{0,1\}$ are amplitude reflection coefficients representing the on/off states of each element, and $\phi_l \in [0, 2 \pi)$ are the different phase shifts, both of which are configurable. Note that the “off” state indicates the incident electromagnetic wave is perfectly absorbed, resulting in no reflection \cite{He20}. For further information on practical implementations, interested readers might refer to \cite{Wei21} and references therein.

\begin{figure}[t]
    \centerline{\includegraphics[scale = 0.6]{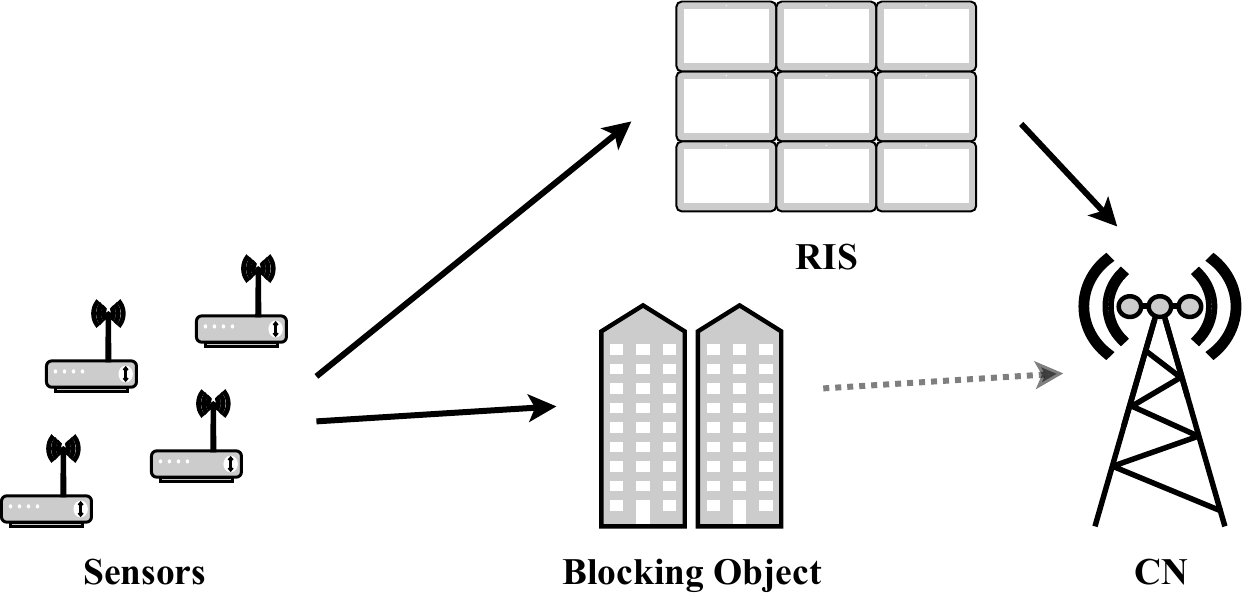}}
    \caption{Illustrative example of a setup with $M = 4$, $L = 9$, and $K = 3$. Solid and dotted lines indicate strong and weak paths, respectively (cf. \cite{Lie23}).}
    \label{fig:1}
\end{figure}

An illustrative example of a setup with $M = 4$ sensors, $L = 9$ reflecting elements, and $K = 3$ antennas is depicted in Fig.~\ref{fig:1}, where the direct link is blocked by a particular object (a critical issue in mmWave and THz bands), and the RIS is utilized to establish an additional path to reach the CN.

The received signal from \eqref{eq:2} can then be written as follows:
\begin{equation}
    \mathbf{y} = \sum_{i = 1}^M \left(\mathbf{q}_i + \mathbf{G}_{\textrm{R}}\bm{\Psi}\mathbf{g}_i\right) s_i + \mathbf{w} = \sum_{i = 1}^M \mathbf{h}_i s_i + \mathbf{w},
    \label{eq:5}
\end{equation}
where $\mathbf{h}_i \in \mathbb{C}^K$ is the effective channel, $\mathbf{G}_{\textrm{R}} \in \mathbb{C}^{K \times L}$ denotes the RIS-CN channel, and $\mathbf{g}_i \in \mathbb{C}^L$ is the channel from sensor $i$ to the RIS. By defining $\bm{\psi} \triangleq [\psi_1,\ldots,\psi_L]^{\textrm{T}} \in \mathbb{C}^L$ as the diagonal entries of $\bm{\Psi}$ (i.e., $\psi_l \triangleq \lambda_l e^{j\phi_l} \in \mathbb{C}$), this expression yields $\mathbf{h}_i = \mathbf{q}_i + \mathbf{G}_{\textrm{C},i} \bm{\psi}$, where $\mathbf{G}_{\textrm{C},i} \triangleq \mathbf{G}_{\textrm{R}}\textrm{diag}(\mathbf{g}_i) \in \mathbb{C}^{K \times L}$ represents the so-called \textit{cascaded channel}.

\section{Methods: Channel estimation} \label{sec:3}
CSI is assumed to be perfectly known at the CN in most scenarios. Typically, this knowledge is acquired via training pilots. However, when considering RIS-aided networks, this procedure presents some drawbacks \cite{DiR20}.

First, because RIS structures are passive, the CSI acquisition must occur at the CN. This involves estimating the cascaded channel of the link between sensors, RIS, and CN, which is non-trivial due to the underdetermination of the system: The unknowns outweigh the available information ($K$ observations for $KL$ channel elements). Second, due to the overhead size required by this operation (i.e., the number of pilot symbols) being proportional to the number of reflecting elements, the estimation becomes prohibitive for large surfaces.

{\color{black} Throughout this work, we consider I-CSI knowledge and aim to design a parameter estimation scheme robust to the corresponding channel estimation errors. For that, we present a feasible protocol for estimating the cascaded link, which overcomes the two previous challenges. In particular, we design the RIS matrix $\bm{\Upsilon}$ during the training phase (distinct from $\bm{\Psi}$, used for data transmission), where the reflection coefficients and phase shifts are selected to minimize the I-CSI errors.} Accordingly, we will consider two different protocols, referred to as \textit{binary} and \textit{non-binary}, both relying on exhaustive searches \cite{Nad20, Wei21}. 

In line with recent findings in the literature \cite{You20, Zhe20}, we later propose a reduction of the resulting overhead by grouping adjacent elements of the RIS and performing joint estimations. Additionally, even with a limited number of pilots, machine learning techniques like those in \cite{Neu18, Zap19}, could further enhance the estimation process (yet such analysis is deferred for future studies).

\begin{figure*}[t]
    \centerline{\includegraphics[scale = 0.8]{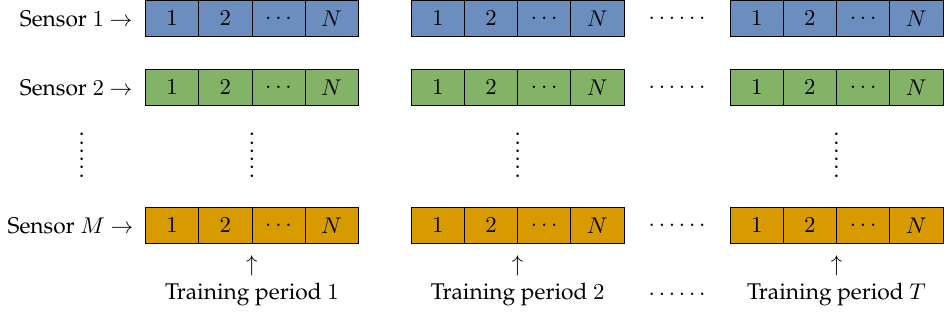}}
    \caption{Training phase of the channel estimation (different colors represent different pilot sequences).}
    \label{fig:2}
\end{figure*}

Each sensor sequentially transmits a training sequence of length $N$ repeated over $T$ training periods towards the CN. As depicted in Fig.~\ref{fig:2}, after the $T$ training periods, we will have $T$ copies of $M$ distinct training sequences of length $N$ (one per sensor). Based on this, at each training period, we tune the response of the RIS so that the CN can compute the channel estimates using the binary and non-binary strategies.

Since this information will be sent sequentially, we will end up with $T$ different received signals (one for each training period). Accordingly, the received signal associated with training period $t \in \{1,\ldots,T\}$ can be expressed as \cite[(7)]{Nad20} 
\begin{equation}
    \mathbf{Y}_t \triangleq \sum_{i = 1}^M \left(\mathbf{q}_i + \mathbf{G}_{\textrm{C},i} \bm{\upsilon}_t \right) \mathbf{p}_i^{\textrm{H}} + \mathbf{W}_t \in \mathbb{C}^{K \times N},
    \label{eq:6}
\end{equation}
where $\mathbf{p}_i = [p_{i,1},\ldots,p_{i, N}] \in \mathbb{C}^N$ is the known uplink (UL) training pilot from sensor $i$ of length $N$ and with $\vert p_{i,m} \vert^2 = P_i$ $\forall m$, and $\mathbf{W}_t \in \mathbb{C}^{K \times N}$ is the AWGN at training period $t$, i.e., $\textrm{vec}(\mathbf{W}_t) \sim\mathcal{CN}(\mathbf{0}_{KN},\sigma_w^2\mathbf{I}_{KN})$. Note that $\bm{\upsilon}_t \in \mathbb{C}^L$ contains the elements of the RIS matrix to be tuned at training period $t$, which depends on the channel estimation approach.

Both channels $\mathbf{q}_i$ and $\mathbf{G}_{\textrm{C}, i}$ can be obtained through conventional MMSE estimation after correlating $\mathbf{Y}_t$ with each pilot. For simplicity, we assume the direct link $\mathbf{q}_i$ is perfectly estimated and removed from the received signal \cite{He20}. Thus, we focus solely on the estimation of the cascaded channel $\mathbf{G}_{\textrm{C}, i}$.

Through pilot correlation\footnote{The training pilots are considered to be mutually orthogonal, ensuring no interference (contamination) is experienced after correlation, i.e., $ \mathbf{p}_i^{\textrm{H}} \mathbf{p}_i  = N P_i$ and $\mathbf{p}_j^{\textrm{H}}\mathbf{p}_i = 0$ for $j \neq i$ \cite{Nad20}. This condition can be satisfied as long as $N \geq M$. Note that for scenarios with an excessively large number of sensors, $M$ could be reduced using device selection techniques (cf. \cite{Lie21}).}, we obtain the sufficient estimation statistic for sensor $i$, i.e., the least-squares estimate of the cascaded channel, at training period $t$ \cite[(1)]{Neu18}:
\begin{equation}
    \mathbf{z}_{i,t} \triangleq \mathbf{Y}_t \frac{\mathbf{p}_i}{\| \mathbf{p}_i \|_2^2} = \mathbf{G}_{\textrm{C},i}\bm{\upsilon}_t + \frac{1}{N P_i} \mathbf{W}_t\mathbf{p}_i  \in \mathbb{C}^K.
    \label{eq:7}
\end{equation}

The set of sufficient statistics $\mathbf{Z}_i = [\mathbf{z}_{i,1},\ldots,\mathbf{z}_{i,T}] \in \mathbb{C}^{K \times T}$ for sensor $i$ can then be written as
\begin{equation}
    \mathbf{Z}_i = \mathbf{G}_{\textrm{C},i}\bm{\Upsilon} + \frac{1}{N P_i} \left[\mathbf{W}_1,\ldots,\mathbf{W}_T\right]\left(\mathbf{I}_T \otimes \mathbf{p}_i\right), 
    \label{eq:8}
\end{equation}
where $\bm{\Upsilon} = [\bm{\upsilon}_1,\ldots,\bm{\upsilon}_T] \in \mathbb{C}^{L \times T}$ is the RIS response matrix during the estimation process, whose design will be discussed in the upcoming subsections.

For notational convenience, we continue with \eqref{eq:8} in vectorized form \cite[(5)]{Bjo10}, i.e., $\mathbf{z}_i \triangleq \textrm{vec}(\mathbf{Z}_i) \in \mathbb{C}^{KT}$:
\begin{equation}
    \mathbf{z}_i = \left(\bm{\Upsilon}^{\textrm{T}} \otimes \mathbf{I}_K \right) \mathbf{g}_{\textrm{C},i} + \frac{1}{N P_i} \left(\left(\mathbf{I}_T \otimes \mathbf{p}_i^{\textrm{T}}\right)\otimes \mathbf{I}_K\right) \textrm{vec}\left(\left[\mathbf{W}_1,\ldots,\mathbf{W}_T\right] \right),     
\label{eq:9}
\end{equation}
where $\mathbf{g}_{\textrm{C},i} \triangleq \textrm{vec}(\mathbf{G}_{\textrm{C},i}) \in \mathbb{C}^{KL}$ represents the cascaded link. From now on, we also assume that the vector $\mathbf{g}_{\textrm{C}, i}$ has a known mean $\bm{\mu}_{\mathbf{g}_{\textrm{C}, i}} \in \mathbb{C}^{KL}$ and covariance matrix $\mathbf{C}_{\mathbf{g}_{\textrm{C}, i}} \in \mathbb{C}^{KL \times KL}$. 

Based on that, the linear MMSE (LMMSE) estimate of the cascaded channel $\mathbf{g}_{\textrm{C}, i}$ is (cf. \cite[(17)]{Zhi23})
\begin{equation}    
    \begin{aligned}
        \hat{\mathbf{g}}_{\textrm{C},i} &\triangleq \underbrace{\mathbf{C}_{\mathbf{g}_{\textrm{C},i}} \mathbf{S}^{\textrm{H}} \left(\mathbf{S} \mathbf{C}_{\mathbf{g}_{\textrm{C},i}} \mathbf{S}^{\textrm{H}} + \frac{\sigma_w^2}{N P_i} \mathbf{I}_{KT} \right)^{-1}}_{\triangleq \mathbf{A}_i} \left(\mathbf{z}_i - \mathbf{S}  \bm{\mu}_{\mathbf{g}_{\textrm{C},i}} \right) + \bm{\mu}_{\mathbf{g}_{\textrm{C},i}} \\
        &= \mathbf{A}_i \mathbf{z}_i + \underbrace{\left(\mathbf{I}_{KL} - \mathbf{A}_i\mathbf{S} \right)\bm{\mu}_{\mathbf{g}_{\textrm{C},i}}}_{\triangleq \mathbf{b}_i} = \mathbf{A}_i \mathbf{z}_i + \mathbf{b}_i \in \mathbb{C}^{KL},    
    \end{aligned}    
    \label{eq:10}    
\end{equation}
where $\mathbf{S} \triangleq \bm{\Upsilon}^{\textrm{T}} \otimes \mathbf{I}_K \in \mathbb{C}^{KT \times KL}$, $\mathbf{A}_i \in \mathbb{C}^{KL \times KT}$, and $\mathbf{b}_i \in \mathbb{C}^{KL}$ are defined for notational brevity. 

We denote $\tilde{\mathbf{g}}_{\textrm{C},i} = \mathbf{g}_{\textrm{C},i} - \hat{\mathbf{g}}_{\textrm{C},i} \in \mathbb{C}^{KL}$ as the (uncorrelated) estimation error with zero mean and covariance matrix $\mathbf{C}_{\tilde{\mathbf{g}}_{\textrm{C},i}} \triangleq \mathbb{E}\left[\tilde{\mathbf{g}}_{\textrm{C},i}\tilde{\mathbf{g}}_{\textrm{C},i}^{\textrm{H}} \vert \mathbf{z}_i \right] = \mathbf{C}_{\mathbf{g}_{\textrm{C},i}} - \mathbf{A}_i \mathbf{S}\mathbf{C}_{\mathbf{g}_{\textrm{C},i}} \in \mathbb{C}^{KL \times KL}$ \cite[Appendix~10A]{Kay93}. As we will see in Section~\ref{sec:4}, this I-CSI knowledge will impact the SIC decoding procedure.

Overall, as discussed in the forthcoming subsections, both strategies (binary and non-binary RIS matrices) entail a trade-off between the channel estimation error and the required overhead training, i.e., smaller values of $T$ yield higher errors but shorter overheads. This will impact the MSE of the parameter estimation (further investigated in Section~\ref{sec:6}).

\subsection{Binary protocol} \label{sec:3.1}
A standard procedure is the so-called \textit{on/off} (or binary) protocol \cite{Wei21}, in which all RIS elements are sequentially turned off except for one (or more) reflector(s). Consequently, the CN will estimate the effective channel(s) corresponding to the active element(s) at each training period. Recall that we distinguish between two setups: \textit{no grouping} and \textit{grouping}.

\subsubsection{No-grouping setup} \label{sec:3.1.1} 
In this case, the RIS response at training period $t$ is directly $\bm{\upsilon}_t = \mathbf{e}_t$, where $\{\mathbf{e}_t\}$ is the set of canonical vectors, i.e., $[\mathbf{e}_t]_t=1$ and $[\mathbf{e}_t]_{t'}=0$ $\forall t' \neq t$. This implies that when the $t$-th element is on (and all the others are off), it is utilized for estimating the $t$-th element of the cascaded channel $\mathbf{g}_{\textrm{C}, i}$. We thus have $\bm{\Upsilon} = \mathbf{I}_T$ with $T = L$.

\subsubsection{Grouping setup} \label{sec:3.1.2}
As the previous procedure involves a prohibitive number of pilots for large surfaces, a feasible alternative is to group adjacent elements and, for each training period, activate the corresponding reflectors simultaneously \cite{You20}. Assuming a group size $G$, the number of training sequences is given by $T = L/G$, and the RIS estimation matrix can then be expressed as $\bm{\Upsilon} \triangleq \mathbf{I}_T \otimes \mathbf{1}_G$, representing the activation of different groups. For instance, if $G = 2$ and $L = 6$, two adjacent elements will be activated simultaneously, resulting in the following matrix:
\begin{equation}
    \bm{\Upsilon} = \begin{bmatrix} 1 & 1 & 0 & 0 & 0 & 0 \\ 0 & 0 & 1 & 1 & 0 & 0 \\ 0 & 0 & 0 & 0 & 1 & 1 \end{bmatrix}^{\textrm{T}}.
    \label{eq:11}
\end{equation}

\subsection{Non-binary protocol} \label{sec:3.2}
Instead of adopting a binary matrix RIS (where $\lambda_l = 0$ is the off state, and $\lambda_l = 1$ the on state), an alternative strategy involves considering all reflectors to be active and designing the phase shifts at each training period \cite{Nad20}. In other words, we define the RIS response as $\bm{\upsilon}_t = [\lambda_{t,1} e^{j\phi_{t,1}},\ldots, \lambda_{t, L} e^{j\phi_{t, L}}]^{\textrm{T}}$ with amplitude coefficients $\lambda_{t,l} = 1$ (activated) and phase shifts $\phi_{t,l} \in [0, 2\pi)$ to estimate the cascaded channel at period $t$. 

Note that this approach applies to both grouping and no-grouping cases, although in the latter we will have $T = L$. As an example, when $G = 1$ (i.e., no grouping), the authors of \cite{Nad20} discussed a discrete Fourier transform-based approach where the elements of $\bm{\Upsilon}$ are directly given by $[\bm{\Upsilon}]_{t,l} = \exp(-\mathrm{i} 2\pi(t-1)(l-1)/L)$. However, the extension to the case with $G > 1$ remains unclear (see Section~\ref{sec:5}).

\section{Methods: Short-packet SIC decoding} \label{sec:4}
As previously mentioned, in this paper, we apply SIC to mitigate the impact of the interference during the communication phase of the devices \cite{Liu22}. In SIC, the different signals are decoded sequentially following a specific order (as discussed in Section~\ref{sec:5}), and, at each step, the previously decoded signals are canceled through subtraction. More precisely, the message from sensor $i$ is first one recovered from the spatial filter output $\mathbf{f}_i^{\textrm{H}}\mathbf{y}$ and its contribution is later removed from the processed signal of the remaining undecoded sensors $\mathbf{f}_j^{\textrm{H}} \mathbf{y}$ (cf. \cite{Dai18}).

Without loss of generality, let us consider an order $\mathbf{o} = [o_1,\ldots,o_M]$ such that $o_i \in \{1,\ldots,M\}$ denotes the step at which sensor $i$ is decoded. The received SINR is \cite{Dai18}:
\begin{equation}
    \rho_i = \frac{P_i \vert \mathbf{f}_i^{\textrm{H}}\mathbf{h}_i \vert^2 }{\displaystyle \sigma_w^2 \| \mathbf{f}_i \|_2^2 + \sum_{j : o_j < o_i} \gamma_j P_j \vert \mathbf{f}_i^{\textrm{H}}\mathbf{h}_j \vert^2 + \sum_{j : o_j > o_i} P_j \vert \mathbf{f}_i^{\textrm{H}}\mathbf{h}_j \vert^2},
    \label{eq:12}
\end{equation}
where $\gamma_j \in \{0,1\}$ is a binary random variable (RV) representing the cancelation failure of the previous signals due to FBL communication errors, i.e., $\gamma_j \sim \textit{Ber}(\textrm{PER}_j)$ with $\textrm{PER}_j$ the packet error rate (PER) or decoding error probability of sensor $j$ \cite{Tse05}. 

However, even with the acquisition of I-CSI, residual power remains due to imperfect cancelation, i.e., 
\begin{equation}
    \rho_i = \frac{P_i \vert \mathbf{f}_i^{\textrm{H}}\hat{\mathbf{h}}_i \vert^2 }{\displaystyle \sigma_w^2 \| \mathbf{f}_i \|_2^2 + \sum_{j : o_j \leq o_i} (1 - \gamma_j)P_j \vert\mathbf{f}_i^{\textrm{H}}\tilde{\mathbf{h}}_j\vert^2 + \sum_{j : o_j < o_i} \gamma_j P_j \vert \mathbf{f}_i^{\textrm{H}}\mathbf{h}_j \vert^2 + \sum_{j : o_j > o_i} P_j \vert \mathbf{f}_i^{\textrm{H}}\mathbf{h}_j \vert^2},
    \label{eq:13}
\end{equation}
where $\hat{\mathbf{h}}_i \triangleq \mathbf{q}_i + \hat{\mathbf{G}}_{\textrm{C}, i}\bm{\psi} \in \mathbb{C}^K$ denotes the estimated effective channel (assuming that the direct channel $\mathbf{q}_i$ is perfectly known or estimated at the CN) and $\tilde{\mathbf{h}}_i \triangleq \mathbf{h}_i - \hat{\mathbf{h}}_i = \tilde{\mathbf{G}}_{\textrm{C},i}\bm{\psi} \in \mathbb{C}^K$ represents the corresponding zero-mean channel estimation error. Here, $\hat{\mathbf{G}}_{\textrm{C},i} \in \mathbb{C}^{K \times L}$ refers to the estimated cascaded channel and $\tilde{\mathbf{G}}_{\textrm{C},i} \in \mathbb{C}^{K \times L}$ denotes the resulting error. Hence, the terms $\vert \mathbf{f}_i^{\textrm{H}} \tilde{\mathbf{h}}_j\vert^2 = \vert \mathbf{f}_i^{\textrm{H}} \tilde{\mathbf{G}}_{\textrm{C}, i} \bm{\psi} \vert^2$ represent the residual powers originating from imperfect channel knowledge (i.e., parts of $\{\mathbf{f}_i^{\textrm{H}}\mathbf{h}_{j:o_j<o_i}\}$ not completely suppressed after decoding).

It can be demonstrated that for the suitable performance of the SIC procedure, it is imperative to have unbalanced received powers to distinguish the signals from different devices (i.e., power-domain NOMA \cite{Isl17}). For instance, in scenarios with equal quality of service requirements, the optimal distribution of received power follows an exponential profile \cite{And03}. However, since our objective is to minimize the MSE when estimating $\bm{\theta}$, the optimal distribution may vary.

In conventional UL wireless networks, the difference in received powers is achieved by adjusting the transmit powers of the sensors or by designing a set of receive spatial filters at the CN (cf. Section~\ref{sec:5}) \cite{And05}. However, in MTC scenarios, power allocation mechanisms at the transmitter side are impractical due to the involvement of a large number of devices. Even if feasible, the simplicity of the sensors limits the number of available power levels and their adjustability. Hence, the set of powers $P_i$ is generally considered to be fixed. In our scenario, the diverse channel gains can be adjusted due to the RIS usage. Thus, the role of the RIS is two-fold: (i) to enhance the quality of the received signals and (ii) to adapt the effective channels to the SIC decoding procedure.

{\color{black} To determine the decoding error probability (as a function of the SINR), we need to define a specific MCS. In this work, to deal with a more general expression, we employ the approximation from \cite[(1)]{Pol10}, which provides an accurate upper bound on the data rate based on Gaussian codewords. In that sense, the corresponding PER is given by:}
\begin{equation}
    \textrm{PER}_i \approx Q\left(\sqrt{n_i} \frac{C(\rho_i) - R_i}{\sqrt{V(\rho_i)}} \right),
    \label{eq:14}
\end{equation}
where $n_i$ is the number of transmit symbols (or packet length), $Q(\cdot)$ is the Gaussian Q-function, $C(x) = \log (1 + x)$ is the Shannon channel capacity, and $V(x) = (1 - (1 + x)^{-2})\log^2 e$ is the so-called \textit{channel dispersion} (which can be interpreted as an additional communication error). Accordingly, $R_i = l_i/n_i$ denotes the data rate (with $l_i$ the number of information bits). We also assume the same number of symbols for all sensors, given by the product of the bandwidth $B_s$ and the packet (or slot) duration $T_s$, i.e., $n_i = n_s = B_s T_s$ $\forall i$. Note that, under the assumption of I-CSI, $C(\rho_i)$ typically represents a lower bound of the actual data rate \cite{Has03}. 

{\color{black} Needless to say, both $R_i$ and $\textrm{PER}_i$ contain the same information (they are indeed connected via \eqref{eq:14}). That is, the evaluation of the two metrics captures the main communication aspects of the scenario. However, as these error probabilities naturally define the transmission failure used in the SINR, we will keep that formulation.}

Unfortunately, due to the presence of I-CSI and decoding errors in \eqref{eq:13}, the resulting SINR is random: It depends on the previously decoded sensors through their RVs $\gamma_j$ with parameter $\textrm{PER}_j$. Since the decoding system is dynamic, statistical dependence exists between the SINRs (i.e., the different Bernoulli RVs $\gamma_j$ are dependent). This makes the joint distribution of the SINRs challenging to compute (the number of required terms grows exponentially with the number of sensors \cite{Dai13}). Hence, we handle tractable statistics in this section by means of a different SIC decoding procedure \cite{Sch20}. 

Suppose sensor $i$ can only be decoded when the previous ones are correctly decoded (but not completely canceled due to I-CSI), i.e., $\gamma_j = 0$ with $j: o_j < o_i$. Based on this (more) conservative scenario, the SINR for each sensor becomes a binary RV equal to
\begin{equation}
    \rho_i \, \vert \underbrace{\{\gamma_j = 0, \forall j :o_j<o_i\}}_{\triangleq \mathcal{S}_i} = \frac{P_i \vert \mathbf{f}_i^{\textrm{H}}\hat{\mathbf{h}}_i \vert^2 }{\displaystyle \sigma_w^2 \| \mathbf{f}_i \|_2^2 + \sum_{j : o_j \leq o_i} P_j \vert \mathbf{f}_i^{\textrm{H}}\tilde{\mathbf{h}}_j \vert^2 + \sum_{j : o_j > o_i} P_j \vert \mathbf{f}_i^{\textrm{H}}\mathbf{h}_j \vert^2},
    \label{eq:15}
\end{equation}
with probability $\textrm{Pr}\{\mathcal{S}_i\}$, and $0$ (indicating no decoding) with probability $1-\textrm{Pr}\{\mathcal{S}_i\}$.

As a result, errors can arise from either the decoding failure of the sensor of interest or from incorrectly decoded packets of the previous sensors (i.e., error propagation). {\color{black} Since these events might not be disjoint (or mutually exclusive), we could use Fréchet’s and Boole’s inequalities to obtain tractable bounds for the PER in \eqref{eq:14} (cf. \cite[(71)]{Sch20}):}
\begin{equation}
    \max\left(\textrm{PER}_i \vert \mathcal{S}_i, \textrm{PER}_{i - 1}\right) \leq \textrm{PER}_i \leq \textrm{PER}_i \vert \mathcal{S}_i + \textrm{PER}_{i - 1},
    \label{eq:16}
\end{equation}
where $\textrm{PER}_i \vert \mathcal{S}_i$ is the PER of sensor $i$ when the previous sensors are perfectly decoded but not completely canceled due to I-CSI errors. This probability is given by the approximation in \eqref{eq:14} with the previous (conditional) SINR $\rho_i \, \vert \, \mathcal{S}_i$ from \eqref{eq:15}. {\color{black} Note that the right-hand side in \eqref{eq:16}, known as the union bound, represents the worst-case scenario and, at the same time, serves as a way to visualize the error propagation. In this work, we also follow a worst-case rationale but represent this phenomenon via the probabilities of the intersections $\mathcal{S}_i$ introduced in \eqref{eq:15}. Hence, as further discussed in Section~\ref{sec:5}, we are more interested in $\textrm{PER}_i \vert \mathcal{S}_i$ than in $\textrm{PER}_i$, yet the union bound above can be interpreted as the connection between both probabilities.}

However, due to the presence of estimation errors $\tilde{\mathbf{h}}_j$, the SINR in \eqref{eq:15} remains random. To address this issue, we adopt the widely used approach from \cite{Has03}, i.e., conditioning on the observations $\mathbf{z} = [\mathbf{z}_1^{\textrm{T}},\ldots,\mathbf{z}_M^{\textrm{T}}]^{\textrm{T}}$ and treating the interference arising from the I-CSI as uncorrelated Gaussian noise. This approach represents a worst-case scenario and, as derived in Appendix~\ref{app:a}, leads to the expression
\begin{equation}
    \rho_i \, \vert \, \mathcal{S}_i,\mathbf{z} = \frac{P_i \vert \mathbf{f}_i^{\textrm{H}} \hat{\mathbf{h}}_i \vert^2 }{\displaystyle \sigma_w^2 \| \mathbf{f}_i \|_2^2 + \sum_j P_j \textrm{tr}\left(\mathbf{C}_{\tilde{\mathbf{g}}_{\textrm{C},j}}\left(\bm{\psi} \bm{\psi}^{\textrm{H}} \otimes \mathbf{f}_i\mathbf{f}_i^{\textrm{H}}\right)\right) + \sum_{j : o_j > o_i} P_j  \vert \mathbf{f}_i^{\textrm{H}}  \hat{\mathbf{h}}_j \vert^2},
    \label{eq:17}
\end{equation}
which cannot be interpreted as a conventional SINR because it involves both instantaneous and average magnitudes. This ratio, often referred to as \textit{effective} SINR \cite{Bjo17}, serves only as a tractable lower bound for the true SINR defined in \eqref{eq:15}, which accurately represents the actual system performance (cf. \cite{Cao21}).

{\color{black} The expression in \eqref{eq:17} represents the error propagation typical of SIC decoding systems. Remarkably, unlike conventional approaches with perfect CSI, at each step of the SIC process, we are only able to suppress the known parts of the previously decoded signal. This means we will accumulate the uncanceled estimation errors along the sequential procedure. These errors are treated as additional uncertainty (or interference), as well as the ones associated with the desired information. For a more detailed proof, please refer to Appendix~\ref{app:a}.}

Finally, note that, since the decoding error probability in \eqref{eq:16} is a function of the SINR in \eqref{eq:17}, it will also be conditioned on $\mathbf{z}$ (i.e., $\textrm{PER}_i \vert \mathcal{S}_i, \mathbf{z}$) and, thus, we must average over its distribution. This will be discussed in the upcoming section.

\section{Methods: Robust system design} \label{sec:5}
This work aims to design the RIS to minimize the parameter estimation error, considering I-CSI. The CN must estimate the cascaded channel before the transmission for that task. Therefore, in this section, in addition to optimizing the matrix $\bm{\Psi}$ for data transmission, we will also include the optimization of the RIS matrix $\bm{\Upsilon}$ (only in the non-binary case) and the group size $G$ for the training phase.

As mentioned in Section~\ref{sec:3}, it is infeasible to assume perfect CSI knowledge, and estimating the channel becomes a challenging task in an RIS-aided environment, especially for large $L$, where the size of the required overhead may be prohibitive. Indeed, the number of pilot symbols $n_p \triangleq N T = N L/G$ and the number of transmitted symbols $n_s$ are limited by the temporal variation of the channel \cite{Mar10}:
\begin{equation}
    n_p + n_s = n_c,
    \label{eq:18}
\end{equation}
with $n_c$ the number of symbols equivalent to the coherence time \cite{Tse05}. This constraint on the physical resources translates to the following trade-off: larger training phases improve the channel estimation errors but also compromise the parameter estimation accuracy as fewer symbols are available for data transmission \cite{Zhi23}. {\color{black} This compromise will be optimized during the system design stage via the group size $G$ (cf. Subsection \ref{sec:3.1.2}). More precisely, the larger $G$ is, the more adjacent RIS elements are gathered to estimate the cascaded channel and, thus, the more symbols are available for data transmission (and vice versa).}

Recall that the decoding error probability in \eqref{eq:14} decays rapidly when increasing the blocklength (in the limit case, we have $\textrm{PER} \to 0$ as $n_s \to \infty$, which corresponds to Shannon's approach). Conversely, when $n_p$ is small, the channel estimation will be poor in exchange for more transmit symbols $n_s$, although the resulting CSI errors might also limit the parameter estimation accuracy. Finding the right balance between $n_p$ and $n_s$ is then crucial for good performance.

Under the decoding procedure described in Section~\ref{sec:4} and for a decoding order $\mathbf{o}$, the MSE (conditioned on $\mathbf{z}$) of the parameter estimation can be written as:
\begin{equation}
    \epsilon \vert \mathbf{z} \triangleq \sum\nolimits_{i = 1}^M \varphi_{i} \epsilon_i \vert \mathbf{z} + \varphi_{0} \epsilon_0 \vert \mathbf{z},
    \label{eq:19}
\end{equation}
where $\epsilon_i \vert \mathbf{z}$ is the “raw” MSE when the information from the sensors in $\{j: o_j \leq o_i\}$ is available (i.e., correctly decoded), and $\varphi_{i} \triangleq \textrm{Pr}\{\gamma_j = 0, \forall j: o_j \leq o_i \, \textrm{and} \, \gamma_{j: o_j = o_i + 1} = 1 \vert \mathbf{z}\}$ is the associated probability (whose analytic closed-form expression is derived in Appendix~\ref{app:b}). Then, $\epsilon_0 \vert \mathbf{z}$ represents the case of complete communication failure, when no sensors are correctly decoded, and $\varphi_{0} \triangleq \textrm{Pr}\{ \gamma_{j:o_j = 1} = 1 \vert \mathbf{z}\} = 1 - \sum_{i = 1}^M \varphi_{i}$.

Using LMMSE estimation for parameter $\bm{\theta}$, the errors $\epsilon_i \vert \mathbf{z}$ follow from \cite[(3.20)]{Bjo17}
\begin{equation}
    \epsilon_i \vert \mathbf{z} \triangleq \textrm{tr}\left(\mathbf{C}_{\bm{\theta}} - \mathbf{C}_{\bm{\theta}}\mathbf{V}_i^{\textrm{T}} \left(\mathbf{V}_i\left(\mathbf{C}_{\bm{\theta}} + \mathbf{C}_{\bm{\eta}}\right)\mathbf{V}_i^{\textrm{T}}\right)^{-1}\mathbf{V}_i\mathbf{C}_{\bm{\theta}}\right),
    \label{eq:20}
\end{equation}
where $\mathbf{V}_i \in \{0,1\}^{i \times M}$ is a binary matrix indicating the correctly decoded messages. 

The overall MSE is then obtained by averaging $\epsilon \vert \mathbf{z}$ w.r.t. the statistics of $\mathbf{z}$, i.e.,
\begin{equation}
    \varepsilon \triangleq \mathbb{E}_{\mathbf{z}}\left[\epsilon \vert \mathbf{z} \right] = \int_{\mathcal{Z}} \epsilon \vert \mathbf{z} f_{\mathbf{z}}(\mathbf{z}) d\mathbf{z},
    \label{eq:21}
\end{equation}
where $f_{\mathbf{z}}(\mathbf{z}): \mathbb{C}^{K T M} \to \mathbb{R}$ is the observations' joint distribution with support $\mathcal{Z}$, which depends on the channel estimation method (cf. Section~\ref{sec:3}). Further discussion on this topic is provided in Section~\ref{sec:6}, where we will concentrate on specific channel models. However, since the conditional error $\epsilon \vert \mathbf{z}$ is combinatorial, the integral in \eqref{eq:21} is solvable only numerically. {\color{black} Note that, unlike \eqref{eq:20}, which only represents the performance in parameter estimation, the MSE in \eqref{eq:21} is averaged over the possible communication failures via the PERs. This means that this last metric also includes the statistics of the errors that might occur during data transmission.}

Depending on the order $\mathbf{o}$, the SIC performance can be affected considerably \cite{And03}. Hence, its impact must be considered in minimizing the average error $\varepsilon$. However, while the spatial filters $\mathbf{f}_i$ also affect the decoding process and the ultimate MSE, this paper's main focus is on the robust RIS design. Optimization of the receive beamformers $\mathbf{f}_i$ will be the subject of future research.

In this study, we generate the different $\mathbf{f}_i$ according to the estimated effective channels $\hat{\mathbf{h}}_i$. Motivated by reasoning in \cite{Mar10, Zhi23}, we employ maximum ratio combining (MRC) as a low-complexity (linear) receive beamforming technique, i.e., $\mathbf{f}_i = \hat{\mathbf{h}}_i$. Since interference is already eliminated with SIC, using MRC is justified as it increases the received signal strength, particularly for large $K$. However, it is worth noting that all the derivations presented in this work are applicable regardless of the linear processing structure used, such as zero-forcing or MMSE, among others. 

As a result, considering the resource constraint in \eqref{eq:18} and the necessity for $N \geq M$ to ensure orthogonality among training pilots (or $N = M$ for simplicity), the joint optimization problem is formulated as:
\begin{equation}
    \{\bm{\Upsilon}^{\star},G^{\star},\bm{\Psi}^{\star},\mathbf{o}^{\star}\} = \textrm{argmin} \, \, \varepsilon \quad \textrm{s.t.} \quad M L/G = n_c - n_s.
    \label{eq:22}
\end{equation}

Due to the expression of $\varepsilon$, \eqref{eq:22} is solved separately using the AO method. In other words, the joint optimization is decomposed into separate sub-problems. Each variable is found while the rest are fixed, and the searches are alternated until convergence is attained \cite{Wu19}. Since, at each iteration, we either maintain the same MSE or improve its value, the convergence of the MSE is always assured. {\color{black} This results in a feasible yet suboptimal approach, as AO does not guarantee obtaining the global optimum. Needless to say, the nonconvex nature of the problem formulated in \eqref{eq:22} prevents any method from attaining a global stationary point (only local optimums are possibly achievable).} For clarity, we dedicate different subsections to each sub-problem.

Note that the proposed scheme implies a low overhead because the optimization is conducted at the CN and depends on statistical magnitudes such as the MSE in \eqref{eq:21}, which averages over the distributions of parameters and channels. The decoding order $\mathbf{o}$ will then be updated only when these statistics vary, and not as frequently as on an instantaneous\footnote{\color{black} If the sensor mobility is low, CSI acquisition would be even simpler and less costly. As discussed in Subsections~\ref{sec:6.1} and \ref{sec:6.2}, by “low mobility” we mean that in the extreme case, the coherence time can be up to $50$ ms \cite{Els17} (which is equivalent to a coherence block of $n_c = 10000$ time-frequency samples for the considered NB-IoT coherence bandwidth of $200$ kHz). Generally, though, we consider far lower values to work with highly dynamic (more realistic) setups, e.g., $n_c = 100$ (which corresponds to speeds of $150$ km/h).} basis. Similarly, the RIS matrices will also require low feedback for configuration \cite{DiR20, Zhi23}. Thus, once the solution to \eqref{eq:22} is found, simple and practical signaling can be easily implemented, further emphasizing the feasibility of our design. {\color{black} This means that any (sudden) increase in the number of sensors or reflecting elements can be handled thanks to this long-term design (scalability is, therefore, not an issue). In particular, as verified in Section 6, we can deal with velocities around $150$ km/h.}

\subsection{RIS matrices} \label{sec:5.1}
For a fixed group size $G$, decoding order $\mathbf{o}$, and RIS estimation matrix $\bm{\Upsilon}$, we have:
\begin{equation}
    \bm{\Psi}^{\star} = \textrm{argmin} \, \, \varepsilon,
    \label{eq:23}
\end{equation}
which can be solved sequentially. In essence, since $\bm{\Psi}$ (or, equivalently, its diagonal $\bm{\psi}$) contains variable elements $\phi_{l} \in [0,2\pi)$, each of them can be found using the bisection method\footnote{\color{black} This allows to extend our design to include non-linearities arising due to mutual coupling (cf. \cite{Rab24}).} when the rest are fixed (cf. \cite{Boy04}). This operation is repeated until convergence is achieved (guaranteed, as each iteration yields the same or a lower MSE). The same procedure can also be used to optimize $\bm{\Upsilon}$ in the case of non-binary channel estimation protocols (cf. Section~\ref{sec:3}), although the search space will increase from $L$ to $TL$ elements. 

{\color{black} Recall that, in the binary case, the reflecting matrix is given by $\bm{\Upsilon} = \mathbf{I}_T \otimes \mathbf{1}_G$ (cf. Subsection~\ref{sec:3.1.2}), meaning the corresponding RIS elements are $\psi_l = 1$ ($\lambda_l = 1$ and $\phi_l = 0$). Hence, the phase-shift optimization discussed here does not apply. At the same time, having fewer degrees of freedom translates into a lower complexity (we make fewer decisions/computations) but at the expense of poorer performance. In that regard though, since our scheme is based on long-term statistical information, we could also afford the longer execution times of the non-binary technique.}

\subsection{Group size} \label{sec:5.2}
To find the optimal group size, problem \eqref{eq:22} gives:
\begin{equation}
    G^{\star} = \textrm{argmin} \, \, \varepsilon \quad \textrm{s.t.} \quad M L/G = n_c - n_s,
    \label{eq:24}
\end{equation}
which is solved by a one-dimensional search since $1 \leq G \leq L$.

\subsection{Decoding order} \label{sec:5.3}
Given $\bm{\Upsilon}$, $G$, and $\bm{\Psi}$, the original problem transforms into:
\begin{equation}
    \{\mathbf{o}^{\star}\} = \textrm{argmin} \, \, \varepsilon,
    \label{eq:25}
\end{equation}
which possesses a combinatorial nature with complexity $\mathcal{O}(M!)$. Thus, an exhaustive search is not feasible (even for a small number of active sensors $M$). Although $M$ can be reduced with device selection techniques, like those outlined in \cite{Lie21}, the computational cost remains unmanageable. Here, we propose heuristic strategies to determine $\mathbf{o}$.

\begin{algorithm}[t]
    \begin{algorithmic}[1]
        \For{$s = 1:M$}
            \State Find next sensor: $j = \underset{j:o_j \geq s}{\textrm{argmin}} \, \, \varepsilon_s$
            \State Add next sensor: $o_j = s$
        \EndFor
    \end{algorithmic}
    \caption{Greedy sensor ordering \cite{Qia19}}
    \label{alg:5.1}
\end{algorithm}

\subsubsection{Received power-based solution} \label{sec:5.3.1}
As the PER increases with the deterioration of the SINR (cf. \eqref{eq:14}), a common approach to determining the decoding order involves assessing the (processed) mean gains of the (estimated) effective channels of the various sensors \cite{Isl17}. Specifically, when decoding the CN first, it may struggle to decode signals from devices with poor channels due to significant interference (without cancelation). Conversely, devices with better conditions can be decoded more easily \cite{Sch20}. Thus, one can consider the vector $\mathbf{o}$ to represent the descending order of the (average) received powers of the effective signals $P_i \mathbf{f}_i^{\textrm{H}}\hat{\mathbf{h}}_i$.

\subsubsection{Measurement-based solution} \label{sec:5.3.2}
The strategy above overlooks the quality of the sensors' observations that affects the individual “raw” MSEs $\epsilon_i \vert \mathbf{z}$ (cf. \eqref{eq:20}). This can be quantified by the ratio $\xi_i \triangleq [\mathbf{C_{\bm{\theta}}}]_{i, i}/[\mathbf{C_{\bm{\eta}}}]_{i, i}$, which resembles a signal-to-noise ratio (SNR). An alternative is then to arrange the order of the devices based on this comparison between parameters and measurement noises. Unlike the previous method, this ordering will affect neither $\bm{\Psi}$, $\bm{\Upsilon}$, nor $G$ ($\xi_i$ remains fixed and independent of the RIS matrices and group size), thereby eliminating the need for updating $\mathbf{o}$ in the AO. {\color{black} Recall that the variances $[\mathbf{C}_{\bm{\eta}}]_{i,i}$ correspond to the power of the measurement noises in \eqref{eq:1}. Hence, the SNR $\xi_i$ can be cast as a parameter estimation-related metric, since no communication aspects, such as quantization/thermal noise, are taken into account.}

\subsubsection{Combined solution} \label{sec:5.3.3}
We can integrate the previous approaches to account for both channel gain and measurement quality. In this subsection, we will first introduce a greedy-based technique similar to the one described in \cite{Qia19}. The core concept is to construct the vector $\mathbf{o}$ sequentially, such that at each stage we select the device that minimizes the estimation error defined in \eqref{eq:21}, incorporating both transmission quality (via PERs) and estimation performance (via the “raw” MSEs $\epsilon_i \vert \mathbf{z}$). 

More precisely, at stage $s \in \{1,\ldots,M\}$, we select the device from the set of  sensors $\{j : o_j > s \}$ that minimizes the average MSE when decoding only the first $s$ messages:
\begin{equation}
    \varepsilon_s = \mathbb{E}_{\mathbf{z}}\left[\sum_{i : o_i \leq s} \varphi_{i} \epsilon_i \vert \mathbf{z} + \varphi_{0} \epsilon_0 \vert \mathbf{z} \right],
    \label{eq:26}
\end{equation}
where $\varphi_{0} =  1 - \sum_{i: o_i \leq s} \varphi_{i}$ now denotes the probability of failure in retrieving the messages from the ordered set of sensors (rather than all devices), which simultaneously represents the probability of not correctly decoding the first sensor (cf. \eqref{eq:19}). Once $s = M$ is reached, the vector $\mathbf{o}$ will be fully determined. This is summarized in Algorithm~\ref{alg:5.1}. 

Alternatively, we can sequentially select the decoding order while considering the complete MSE $\varepsilon$. To achieve this, we begin by initializing $\mathbf{o}$ randomly, and at each position, we explore all available possibilities. At stage $s$, we eliminate the sensors decoded up to stage $s - 1$ and select the best one from among the set of remaining devices $\{j: o_j \geq s\}$, meaning the sensor that minimizes $\varepsilon$ and has not yet been decoded. For example, if $M = 3$ and the initial order is $\mathbf{o} = [2,3,1]$, we can choose between sensors $1$, $2$, and $3$ for the first stage (where no device has been decoded). 

Accordingly, the positions of the involved devices must be swapped. For instance, if $2$ is the selected terminal, the order will become $\mathbf{o} = [2,1,3]$. Consequently, at stage $2$, only devices $1$ and $3$ are available. Ultimately, if we choose sensor $3$, the sole remaining option at the last stage will be terminal $1$, resulting in the ultimate decoding order $\mathbf{o} = [3,1,2]$. This process is elucidated in Algorithm~\ref{alg:5.2}.

\begin{algorithm}[t]
    \begin{algorithmic}[1]
        \State Initialize $\mathbf{o}$
        \For{$s = 1:M$}
            \State Find next sensor: $j = \underset{j : o_j \geq s}{\textrm{argmin}} \, \, \varepsilon$
            \State Update order: $[o_j,o_{j'}] = [s,s']$ s.t. $o_{j'} = s$ and $o_j = s'$
        \EndFor
    \end{algorithmic}
    \caption{Combined sensor ordering}
    \label{alg:5.2}
\end{algorithm}

\section{Results and discussion} \label{sec:6}
This section examines the average MSE $\varepsilon$ (after RIS and SIC decoding order optimization) across various setups. We employ a realistic MTC network, adhering to the parameters and guidelines outlined in the 3GPP and ITU standards \cite{3GPP38214, ITU09}. 

Regarding the actual scenario under evaluation, we presume that the measurements originate from the database compiled by the Intel Berkeley Research Lab, incorporating up to $M = 20$ active sensing devices\footnote{\color{black} Recall that orthogonal allocation becomes infeasible for a sufficiently large $M$ since the total number of symbols $n_t$ (equal to $n_c$ in NB-IoT) must be distributed among the sensors. This can be even more critical in high-mobility scenarios, where $n_c$ is rather small. For example, for $n_c = 100$ (and $M = 20$), only $5$ time-frequency samples are given to each device. Because of this, the resulting PERs would be too large for reliable communication.} situated within a confined research area \cite{Bod04}. {\color{black} Accordingly, the sources of uncertainty in the observations (or measurement noise) are determined by such indoor deployment, e.g., sensors’ hardware and surroundings (among others). Needless to say, our proposed scheme is not specific to this setup and can be applied to other environments (urban, rural, etc.).

Notably, given the sporadic nature of MTC transmissions, for instance, for periodic reports of $l = 10$ kb every $t = 6$ h at $r = 25$ kbps, having $M = 20$ active sensors is equivalent to supporting a total of $(rt/l)M \approx 10^6$ terminals over the deployment area (i.e., mMTC). This means our approach holds even for large networks.
}

Subsequently, we dedicate an initial subsection to discuss these practical issues. 

\subsection{Practical issues} \label{sec:6.1}
Throughout the simulations, we adopt the micro-urban scenario outlined in \cite{ITU09}, where $P_i \in [-10,10]$ dBm $\forall i$, $\sigma_w^2 = N_o B_s$, $N_o = -174$ dBm/Hz, $B_s = 180$ kHz (cf. NB-IoT \cite{3GPP45820}), and a 5G-compliant carrier frequency of $3$ GHz \cite{Bjo22}. We assume a coherence block of $n_c$ OFDM symbols with sub-carrier spacing of $\Delta f = 15$ kHz and of duration $T_o \approx 71.4$ $\mu\textrm{s}$ (equal to $1/\Delta f$ plus the guard intervals), occupying a total coherence bandwidth $B_c = 200$ kHz (equal to $B_s$ plus the guard bands) \cite{Mar10,3GPP38214,Nok16}.

Note that $n_c = T_c B_c$, with $T_c$ the coherence time, is contingent upon the sensors' mobility\footnote{\color{black} In general, we will consider high mobility or rapidly fluctuating channel conditions. To do so, we will set $n_c = 100$ in most experiments. For the coherence bandwidth of $200$ kHz \cite{3GPP45820} and the carrier frequency of $3$ GHz \cite{Bjo22}, this is equivalent to $T_c = 0.5$ ms and maximum speeds of $\approx 150$ km/h.}, thereby conditioning the tuple $n_s$ and $n_p$ (cf. \eqref{eq:18}). The packet duration $T_s = n_s T_o$ will vary based on the group size $G$ or the number of training pilots $n_p$. The individual data rates $R_i = l_i/n_s$ in \eqref{eq:14} are obtained from Table~$5.2.2.1$ in \cite{3GPP38214}, and the number of information bits $l_i$ will also be adjusted to fulfill all temporal requirements, ultimately affecting parameter estimation accuracy.

For the sensors-CN channels, we assume a non-LoS (NLoS) propagation \cite{Zhi23} and a power-law path loss:
\begin{equation}
    \mathbf{q}_i = d_i^{-\alpha_1/2} \bm{\chi}_i,
    \label{eq:27}
\end{equation}
where $d_i$ represents the distance from sensor $i$ to the CN, $\bm{\chi}_i \sim \mathcal{CN} (\mathbf{0}_K,\mathbf{I}_K)$ comprises the set of Rayleigh fading coefficients, and $\alpha_1 = 3.8$ is the decay exponent \cite{Mar10}.

{\color{black} Conversely, given the indoor deployment of the scenario under evaluation (i.e., the Intel Berkeley Research Lab \cite{Bod04}), the channels between the sensors and the RIS can be described by the well-known Rician model \cite{For07, Bjo17}:}
\begin{equation}
    \mathbf{g}_{i} = \delta_{i}^{-\alpha_2/2} \left(\sqrt{\frac{1}{1+F_1}} \bm{\tau}_{i} + \sqrt{\frac{F_1}{1+F_1}} \mathbf{v}(\vartheta_i) \right),
    \label{eq:28}
\end{equation}
where $\delta_i$ is the distance from sensor $i$ to the RIS, $\alpha_2 = 2.2$ denotes the path-loss factor \cite{Cao21}, $F_1 = 3$ dB is the Rician factor\footnote{\color{black} Note that our approach is not specific to particular channels, but any model could be considered. For instance, for Rayleigh fading, one could simply set $F_1$ to zero.}, $\bm{\tau}_{i} \sim \mathcal{CN} (\mathbf{0}_L,\mathbf{I}_L)$ contains the NLoS components, $\mathbf{v}(\cdot) \in \mathbb{C}^L$ is the RIS steering vector, and $\vartheta_i$ is the angle of arrival (AoA) to the RIS.

Likewise, the RIS-CN channel can be written as \cite[(30)]{Guo20}:
\begin{equation}
    \mathbf{G}_{\textrm{R}} = \delta_{\textrm{R}}^{-\alpha_2/2} \left(\sqrt{\frac{1}{1+F_2}} \bm{\Delta}_{\textrm{R}} + \sqrt{\frac{F_2}{1+F_2}} \mathbf{u}(\vartheta_{\textrm{R},1}) \mathbf{v}(\vartheta_{\textrm{R},2})^{\textrm{H}} \right),
    \label{eq:29}
\end{equation}
where $\delta_{\textrm{R}}$ is the distance from the RIS to the CN, $F_2 = 10$ dB denotes the Rician factor, $[\bm{\Delta}_{\textrm{R}}]_{k,l} \sim \mathcal{CN} (0,1)$ $\forall k,l$ represents the set of NLoS components (independent of all $\bm{\tau}_i$), $\mathbf{u}(\cdot) \in \mathbb{C}^K$ is the CN steering vector, $\vartheta_{\textrm{R},1}$ is the AoA to the CN, and $\vartheta_{\textrm{R},2}$ is the angle of departure from the RIS.

Consequently, the observations follow the distribution
\begin{equation}
    \mathbf{z} \sim \mathcal{CN}\left(\begin{bmatrix} \bm{\mu}_{\mathbf{z}_1} \\ \vdots \\  \bm{\mu}_{\mathbf{z}_M} \end{bmatrix}, \begin{bmatrix} \mathbf{C}_{\mathbf{z}_1} & \ldots & \mathbf{C}_{\mathbf{z}_1 \mathbf{z}_M} \\ \vdots & \ddots & \vdots \\  \mathbf{C}_{\mathbf{z}_M \mathbf{z}_1} & \ldots & \mathbf{C}_{\mathbf{z}_M} \end{bmatrix}\right),
    \label{eq:30}
\end{equation}
where the statistical moments $\bm{\mu}_{\mathbf{z}_i} \triangleq \mathbb{E}[\mathbf{z}_i]$, $\mathbf{C}_{\mathbf{z}_i} \triangleq \mathbb{E}[(\mathbf{z}_i - \bm{\mu}_{\mathbf{z}_i}) (\mathbf{z}_i - \bm{\mu}_{\mathbf{z}_i})^{\textrm{H}}]$, and $\mathbf{C}_{\mathbf{z}_i,\mathbf{z}_j} \triangleq \mathbb{E}[(\mathbf{z}_i - \bm{\mu}_{\mathbf{z}_i}) (\mathbf{z}_j - \bm{\mu}_{\mathbf{z}_j})^{\textrm{H}}]$ are determined by the channel estimation strategy (cf. \eqref{eq:10}):
\begin{align}
    \label{eq:31} \bm{\mu}_{\mathbf{z}_i} &= \sqrt{\frac{F_1F_2 \left(\delta_i\delta_{\textrm{R}}\right)^{-\alpha_2}}{(1 + F_1)(1 + F_2)}} \mathbf{S}\textrm{vec}\left(\mathbf{u}(\vartheta_{\textrm{R},1}) \mathbf{v}(\vartheta_{\textrm{R},2})^{\textrm{H}}\textrm{diag}\left(\mathbf{v}(\vartheta_i)\right)\right), \\
    \label{eq:32} \mathbf{C}_{\mathbf{z}_i} &= \frac{\left(\delta_i\delta_{\textrm{R}}\right)^{-\alpha_2}}{(1 + F_1)(1 + F_2)} \mathbf{S}\mathbf{C}_{\mathbf{g}_{\textrm{C},i}}\mathbf{S}^{\textrm{H}} + \frac{\sigma_w^2}{N P_i} \mathbf{I}_{KT}, \\
    \label{eq:33} \mathbf{C}_{\mathbf{z}_i,\mathbf{z}_j} &= \frac{F_1 \delta_{\textrm{R}}^{-\alpha_2} \left(\delta_i\delta_j\right)^{-\alpha_2/2}}{(1 + F_1)(1 + F_2)} \mathbf{S} \left(\textrm{diag}(\mathbf{v}(\vartheta_i) \odot \mathbf{v}(\vartheta_j)^{\textrm{H}}) \otimes \mathbf{I}_K\right)\mathbf{S}^{\textrm{H}},
\end{align}
with $\mathbf{C}_{\mathbf{g}_{\textrm{C}, i}} = ((1 + F_1)\mathbf{I}_{KL} + F_2(\mathbf{I}_L \otimes \mathbf{u}_{\textrm{R}}\mathbf{u}_{\textrm{R}}^{\textrm{H}}))$ the covariance matrix of the cascaded channel $\mathbf{g}_{\textrm{C}, i}$ (cf. \eqref{eq:9}). Similarly, by specifying the expression $\bm{\mu}_{\mathbf{z}_i}$ for $\bm{\Upsilon} = \mathbf{I}_L$, i.e., no RIS estimation matrix, we can derive the mean $\bm{\mu}_{\mathbf{g}_{\textrm{C}, i}}$. All this is proven in Appendix~\ref{app:c}.

In all experiments, the sensors' spatial distribution adheres to the deployment outlined in \cite{Bod04}. We assume the RIS is positioned at the center of the laboratory, equipped with a uniform planar array of $L$ elements. The reflection coefficients $\lambda_l$ in the data transmission matrix $\bm{\Psi}$ are set to $1$ \cite{Wan21}. We also assume the CN is situated at the corner of the research area, equipped with a uniform linear array of $K$ elements.
 
\subsection{Robust system design} \label{sec:6.2}
In this subsection, we present several results to evaluate the performance of our parameter estimation approach. 

Specifically, we assess the average error $\varepsilon$ from \eqref{eq:21} concerning the number of reflecting elements $L$, the number of sensors $M$, the number of CN antennas $K$, and the number of coherence symbols $n_c$. This evaluation spans different setups, where we analyze the channel estimation protocols and SIC decoding orders $\mathbf{o}$ from Sections~\ref{sec:3} and \ref{sec:4}, respectively. For clarity, we show the normalized MSE (NMSE) as $\varepsilon/\textrm{tr}(\mathbf{C}_{\bm{\theta}})$.

\begin{figure}[t]
    \centerline{\includegraphics[scale = 1]{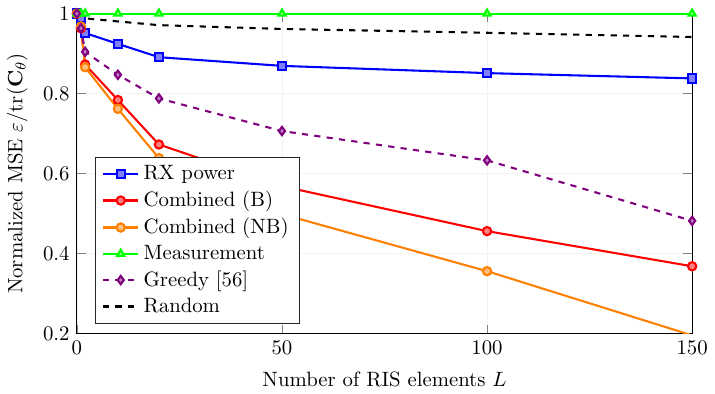}}   
	\caption{NMSE vs. $L$. Solid and dashed lines indicate the proposals and benchmarks, respectively.}    
	\label{fig:3}
\end{figure}

First, we concentrate on the estimation accuracy concerning the size of the surface $L$. This is depicted in Fig.~\ref{fig:3} for $M = 20$, $K = 10$, $n_c = 100$, and the orders described in Subsection~\ref{sec:5.3}. For a broader comparison, we also include the case of a random device ordering. To avoid redundancy and streamline the presentation, the non-binary protocol (NB) is depicted only with the combined solution, while the remaining curves are computed using the binary approach (B).

In most cases, the NMSE decreases with the increase in the number of RIS elements because the surface gain can enhance the SINR. Nevertheless, as discussed in Section~\ref{sec:5}, the performance of the SIC architecture significantly depends on the decoding order. For instance, in the measurement-based strategy, sensors are ordered regardless of channel quality, potentially leading to high communication errors. In brief, the chosen order may yield large PERs\footnote{\color{black} As mentioned in Section~\ref{sec:4}, for a given PER, we can retrieve the corresponding data rate (cf. \eqref{eq:14}). Therefore, if the PERs tend to be one, the data rate will approach zero (i.e., complete system failure).}; that is, devices with poor channel quality might be decoded first, hindering any potential decoding at the CN. Consequently, an almost complete estimation error (NMSE close to $1$) is observed.

In line with this, the combined decoding order utilizing both binary and non-binary protocols outperforms both the received (RX) power-based method and that of \cite{Qia19}. Also, for large $L$, the non-binary approach yields superior results compared to the binary case. These simulations underscore the importance of selecting an appropriate order and RIS training matrix.

\begin{figure}[t]
    \centerline{\includegraphics[scale = 1]{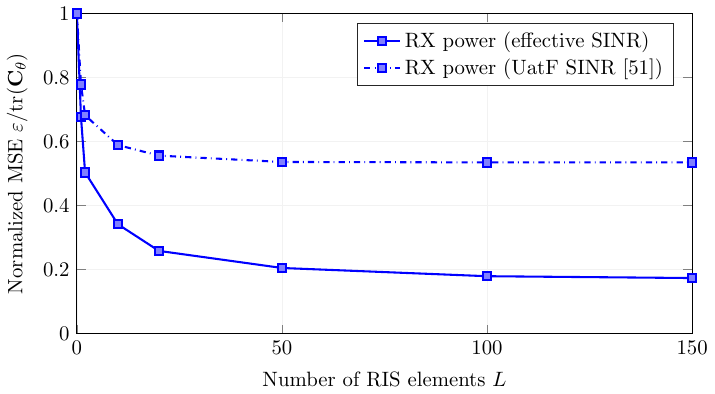}}
    \caption{NMSE vs. $L$ for different SINR bounds.}    
	\label{fig:4}        
\end{figure}

Second, as shown in Fig.~\ref{fig:4}, we compare the NMSE obtained with the effective SINR and the \textit{use-and-then-forget} (UatF) SINR \cite{Bjo17} vs. $L$. The former refers to the expression already derived in \eqref{eq:17} using the bound from \cite{Has03}, whereas the latter corresponds to that from \cite{Med00} described in Appendix~\ref{app:d}. This is conducted for $M = 5$, $K = 10$, $n_c = 100$, and the RX power-based order. Unlike \cite{Has03}, the UatF bound entails poorer channel knowledge (only the mean is known), resulting in higher channel estimation errors and a noticeable performance gap. Given that this behavior is more limiting for larger values of $M$, the NMSE considering \cite{Med00} rapidly tends to $1$.

\begin{figure}[t]
    \centerline{\includegraphics[scale = 1]{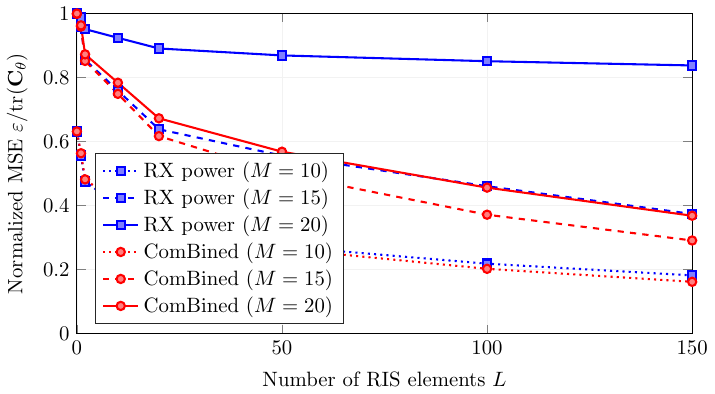}}
    \caption{NMSE vs. $L$ for different $M$.}
    \label{fig:5}
\end{figure}

\begin{figure}[t]
    \centerline{\includegraphics[scale = 1]{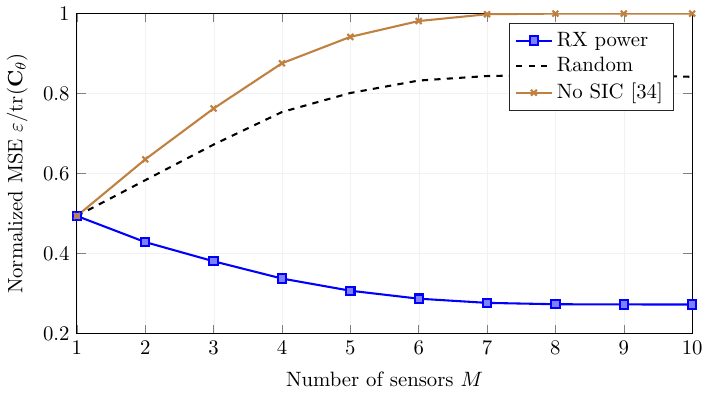}}
    \caption{NMSE vs. $M$.}	
    \label{fig:6}
\end{figure}

The impact of the number of sensors is indeed studied in Fig.~\ref{fig:5}. The RX power and combined (B) (denoted by \textit{comBined}) orders are plotted w.r.t. $L$ for $K = 10$, $n_c = 100$, and different $M$. Similar to previous observations, the accuracy increases with the size of the RIS, yet minor errors are obtained when decreasing the number of MTC terminals. This is unsurprising, as less interference is experienced, thus allowing more messages to be retrieved at the CN. However, this also implies that the overall amount of transmitted information will be lowered. Additionally, for low $M$, capitalizing on the channel conditions also converges to the best results. This suggests that the quality of the measurements starts to play a role in the decoding order in denser scenarios (or, equivalently, in highly correlated environments), which is typical of MTC.

The accuracy in terms of the number of sensors is also illustrated in Fig.~\ref{fig:6} for $L = 50$, $K = 10$, and $n_c = 100$, where we incorporate the case of no SIC decoding (calculated following the derivations in \cite{Lie21}). For a fair comparison, we show the attempts in which only $1$ incorrectly decoded packet is allowed; in other words, communication fails whenever more than $1$ message is unsuccessfully recovered. It is evident that this method performs poorly when $M$ increases (the NMSE tends to $1$). Because interference is not suppressed, the resulting decoding error probabilities are too high for reliable communication. This underscores the necessity of employing interference cancelation methods to avoid the saturation of the NMSE. However, a decreasing trend is observed only when the SIC decoding order\footnote{\color{black} It is important to highlight that, the SIC decoding order must be updated whenever there is an environmental change that modifies the statistics of the scenario (e.g., an increase in the number of sensors or a replacement of devices). However, if that information does not change, the design remains unaltered (is valid regardless of the instantaneous values). The same happens to the figure of merit: the objective function in \eqref{eq:21} varies only with the statistical information ($\varepsilon$ represents the average MSE).} is appropriately chosen (cf. Fig.~\ref{fig:3}). For brevity, only the random and RX power-based selection methods are included for comparison purposes.

\begin{figure}[t]
    \centerline{\includegraphics[scale = 1]{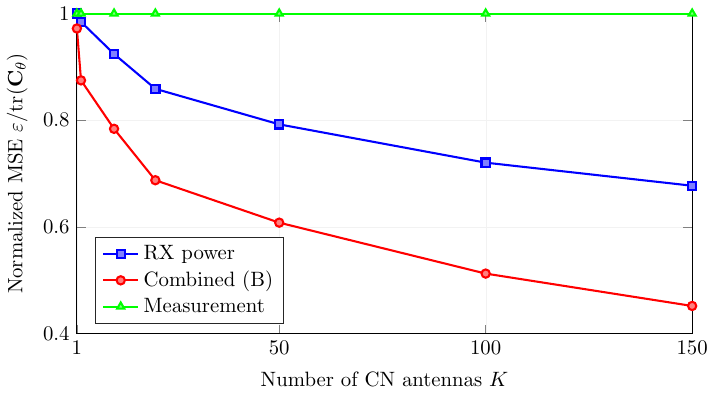}}
    \caption{NMSE vs. $K$.}	
    \label{fig:7}
\end{figure}

To explore the performance when varying the number of CN antennas, we present in Fig.~\ref{fig:7} the NMSE vs. $K$ for $L = 10$, $M = 20$, $n_c = 100$, and the RX power-based and comBined decoding orders. As anticipated, larger antenna arrays yield lower values of $\varepsilon$. In fact, this beamforming gain closely resembles that of the RIS (cf. Fig.~\ref{fig:3}). Thus, as increasing the number of passive elements is cost-effective and more energy-efficient, RISs can be a potential solution for green and sustainable MTC.

\begin{figure}[t]
    \centerline{\includegraphics[scale = 1]{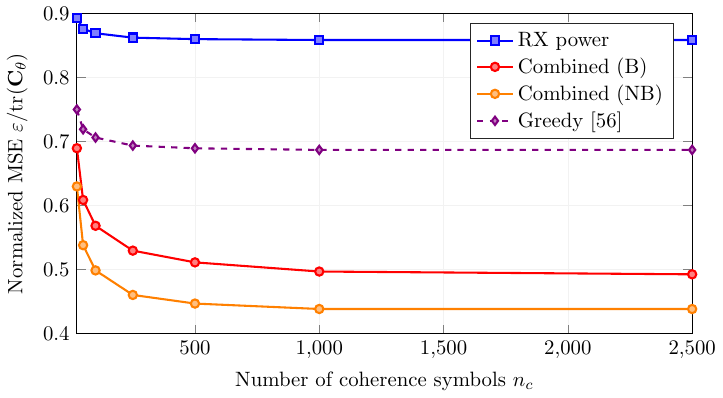}} 
	\caption{NMSE vs. $n_c$. Solid and dashed lines indicate the proposals and benchmarks, respectively.}    
	\label{fig:8}
\end{figure}

In Fig.~\ref{fig:8}, we show the NMSE w.r.t. the number of coherence symbols for $L = 50$, $M = 20$, and $K = 10$. Recall that $n_c$ is the product of the coherence bandwidth $B_c$ and coherence time $T_c$. {\color{black} In low mobility scenarios, a $T_c$ of up to $50$ ms might be considered (e.g., EPA-5 standard \cite{Els17}), which corresponds to $n_c = 10^4$ symbols for the coherence bandwidth of $200$ kHz. However, in this illustration, only up to $2500$ time-frequency samples are shown, given the performance convergence of all curves.} As observed, larger coherence blocks result in smaller errors, as more resources are allocated to channel knowledge and parameter estimation. In other words, more training pilots enable better CSI acquisition\footnote{\color{black} Note that, however, the designed system is more robust w.r.t. I-CSI uncertainty than communication errors (specifically w.r.t. the number of transmit symbols). That is, if $n_s$ is too small, the SINRs become “virtually” zero, and, thus, the impact of channel estimation errors is negligible. In other words, data transmission has more priority than training in NB-IoT environments with high mobility.} while PERs decrease with the number of transmit symbols\footnote{\color{black} The same applies to the data rate, i.e., larger coherence blocks result in a high rate $R_i$ approaching Shannon’s capacity $C(\rho_i)$. Contrarily, shorter coherence times limit the number of transmit symbols and, thus, compromise both PER and data rate.} (cf. \eqref{eq:14}). Once again, the non-binary (NB) protocol with the combined decoding order outperforms the rest of the schemes. For the sake of clarity, the random and measurement-based selection have been omitted. 

{\color{black} As expected, the proposed approaches become sensitive when fewer than $n_c < 250$ symbols are used. Interestingly, though, parameter estimation is really infeasible (NMSE surpassing $0.5$) only for extremely high-speed varying channels, i.e., $n_c < 100$ (or, equivalently, coherence times smaller than $0.5$ ms and velocities above $150$ km/h). 

\noindent
In those regimes, the resulting PERs are too high for reliable communication. However, even in those cases, the combined decoding order (both with binary and non-binary RIS channel estimation matrices) still leads to better results when compared to received power (RX)-based selection or the greedy technique from \cite{Qia19}.}

Finally, to quantify the computational complexity, Table~\ref{tab:1} presents the average execution time of the decoding order optimization w.r.t. $M$. For a comprehensive comparison, the UatF bound is also included (albeit the random and measurement-based selections are omitted, to avoid redundancy). It is evident that while offering better performance (cf. Figs~\ref{fig:3} and \ref{fig:8}), our comBined approach entails a slightly longer execution time than the greedy procedure from \cite{Qia19}. {\color{black} However, since the actual difference is practically null, one could safely argue that their computational complexity is (almost) the same. This can indeed be reasoned by the fact that,  for each iteration $s$ of the sequential procedures (cf. Algorithms~\ref{alg:5.1} and ~\ref{alg:5.2}), both techniques need $M - s + 1$ trials to find the best sensor for the decoding order. This means that, per iteration, the complexity grows linearly (and equally) with that number of search elements. In that sense, the results in Table~\ref{tab:1} also suggest that the iterations required in both cases must be (practically) equal.} Unsurprisingly, and unlike the considered effective (eff.) SINR in \eqref{eq:17}, the use of the looser SINR bound from \cite{Med00} requires fewer operations (as it relies exclusively on CDI; thus, no conditioning is needed).

\begin{table*}[h]
    \caption{Average execution time (in seconds) vs. $M$. All simulated values were obtained using MATLAB R2023b code, implemented on a Windows 10 x64 machine equipped with 64 GB of RAM and an AMD Ryzen 9 5950X 16-Core CPU running at 4 GHz.}
    \begin{center}
    \resizebox{\columnwidth}{!}{        
        \begin{tabular}{|c|c|c|c|c|}
        \hline
        Number of sensors $M$ & RX power (UatF SINR) & RX power (eff. SINR) & Greedy \cite{Qia19} (eff. SINR) & ComBined (eff. SINR) \\ 
        \hline
        \hline
        $10$ & $0.20$ & $1.79$ & $1.83$ & $1.86$  \\
        \hline
        $15$ & $0.42$ & $3.88$ & $3.94$ & $4.06$ \\
        \hline
        $20$ & $0.63$ & $6.27$ & $6.56$ & $6.98$ \\
        \hline
        \end{tabular}                
    }
    \end{center}
    \label{tab:1}
\end{table*}

\section{Conclusions} \label{sec:7}
In this paper, we have addressed the problem of designing an RIS to support parameter estimation in an MTC network. In a setup where single-antenna sensors transmit their correlated measurements to a serving multiple-antenna CN on a NOMA basis, we have presented an estimation scheme relying on the MMSE criterion. The resulting estimate is derived taking into consideration FBL communication and I-CSI errors. To overcome the lack of perfect channel knowledge, we have also studied different methods for acquiring that information feasibly. To mitigate interference among devices, we have introduced the use of SIC and optimized the decoding order. Based on that, the RIS is optimized to minimize the average MSE under channel variation (i.e., coherence time) constraints.

To assess the performance of our approach, we have conducted simulations with actual databases. Results have shown that incorporating RISs in MTC systems can significantly reduce the MSE, especially with a large number of elements. However, parameter estimation accuracy can be compromised when considering fast varying channels. Finally, using SIC decoding becomes essential for achieving good ultimate performance, provided that the decoding order is duly chosen.

\begin{appendices}
\section{Effective SINR} \label{app:a}
Let us define the following signal model to derive the expression in \eqref{eq:17}. Considering the (ultimate) decoding procedure from Section~\ref{sec:4}, the equivalent signal of sensor $i$ is
\begin{equation}
    \hat{s}_i \triangleq \mathbf{f}_i^{\textrm{H}}\hat{\mathbf{h}}_i s_i + \underbrace{\vphantom{\mathbf{f}_i^{\textrm{H}}\sum_{\substack{j=1 \\ o_j < o_i}}^{M}\tilde{\mathbf{h}}_j s_j + \mathbf{f}_i^{\textrm{H}}\sum_{\substack{j=1 \\ o_j > o_i}}^M \mathbf{h}_j s_j + \mathbf{f}_i^{\textrm{H}}\mathbf{w}} \mathbf{f}_i^{\textrm{H}}\tilde{\mathbf{h}}_i s_i}_{\triangleq w_{\textrm{est},i}} + \underbrace{\mathbf{f}_i^{\textrm{H}}\sum_{\substack{j=1 \\ o_j < o_i}}^{M}\tilde{\mathbf{h}}_j s_j + \mathbf{f}_i^{\textrm{H}}\sum_{\substack{j=1 \\ o_j > o_i}}^M \mathbf{h}_j s_j}_{\triangleq w_{\textrm{eff},i}} + \mathbf{f}_i^{\textrm{H}}\mathbf{w},
    \label{eq:34}
\end{equation}
where $\mathbf{f}_i^{\textrm{H}}\hat{\mathbf{h}}_is_i$ is now the desired signal and $w_{\textrm{est},i}$ is treated as an additional (estimation) noise. In that sense, following the worst-case scenario described in \cite{Has03}, the estimation noise $w_{\textrm{est},i}$ and the effective noise $w_{\textrm{eff},i}$ are considered to be uncorrelated and Gaussian distributed with variances $\textrm{var}(w_{\textrm{est},i})$ and $\textrm{var}(w_{\textrm{eff},i})$, respectively. 

As a result, when conditioning on the observation $\mathbf{z}$, the SINR can be expressed as:
\begin{equation}
    \rho_i \, \vert \, \mathcal{S}_i, \mathbf{z} = \frac{\textrm{var}\left(\mathbf{f}_i^{\textrm{H}}\hat{\mathbf{h}}_is_i\vert \mathbf{z}\right)}{\displaystyle \textrm{var}\left(w_{\textrm{est},i}\vert \mathbf{z}\right) + \textrm{var}\left(w_{\textrm{eff},i}\vert \mathbf{z}\right) + \sigma_w^2\left\| \mathbf{f}_i \right\|_2^2},
    \label{eq:35}
\end{equation}
where, due to the MMSE channel estimation, $\hat{\mathbf{h}}_i$ become deterministic functions of $\mathbf{z}$. Additionally, since the spatial filters $\mathbf{f}_i$ are functions of the estimated channels, they are also constant. Hence, given that the transmit signals are independent of $\mathbf{z}$, have zero mean, and power $P_i$, we have:
\begin{equation}
    \textrm{var}(\mathbf{f}_i^{\textrm{H}}\hat{\mathbf{h}}_is_i\vert \mathbf{z}) \triangleq \mathbb{E}\left[\vert \mathbf{f}_i^{\textrm{H}}\hat{\mathbf{h}}_is_i \vert^2 \vert \mathbf{z}\right] = P_i \vert \mathbf{f}_i^{\textrm{H}}\hat{\mathbf{h}}_i \vert^2.
    \label{eq:36}
\end{equation}

Since errors $\mathbf{f}_i^{\textrm{H}}\tilde{\mathbf{h}}_j = \mathbf{f}_i^{\textrm{H}}\tilde{\mathbf{G}}_{\textrm{C},j}\bm{\psi}$ are zero mean with variance:
\begin{equation}    
    \kappa_{i,j} \triangleq \mathbb{E}\left[\mathbf{f}_i^{\textrm{H}}\tilde{\mathbf{h}}_j \tilde{\mathbf{h}}_j^{\textrm{H}}\mathbf{f}_i \vert \mathbf{z}\right] = \textrm{tr}\left(\mathbf{C}_{\tilde{\mathbf{g}}_{\textrm{C},j}}\left(\bm{\psi} \bm{\psi}^{\textrm{H}} \otimes \mathbf{f}_i\mathbf{f}_i^{\textrm{H}}\right)\right),
    \label{eq:37}
\end{equation}
the other second-order moments in \eqref{eq:35} result in:
\begin{align}
    \label{eq:38} \textrm{var}\left(w_{\textrm{est},i}\vert \mathbf{z}\right) &= P_i \kappa_{i,i}, \\
    \label{eq:39} \textrm{var}\left(w_{\textrm{eff},i}\vert \mathbf{z}\right) &= \sum_{j : o_j < o_i} P_j \kappa_{i,j} + \sum_{j : o_j > o_i} P_j (\vert\mathbf{f}_i^{\textrm{H}} \hat{\mathbf{h}}_j\vert^2 + \kappa_{i,j}),
\end{align}
where the terms $\mathbb{E}[\vert \mathbf{f}_i^{\textrm{H}} \mathbf{h}_j s_j \vert^2 \vert \mathbf{z}]$ become $P_j (\vert\mathbf{f}_i^{\textrm{H}}\hat{\mathbf{h}}_j\vert^2 + \kappa_{i,j})$ as the cross products disappear, i.e.:
\begin{equation}
    \mathbb{E}\left[\mathbf{f}_i^{\textrm{H}}\hat{\mathbf{h}}_j \tilde{\mathbf{h}}_j^{\textrm{H}} \mathbf{f}_i \vert s_j \vert^2 \vert \mathbf{z} \right] = P_j \mathbf{f}_i^{\textrm{H}}\hat{\mathbf{h}}_j \mathbb{E}\left[\tilde{\mathbf{h}}_j^{\textrm{H}} \vert \mathbf{z} \right] \mathbf{f}_i = 0.        
\label{eq:40}
\end{equation}

Lastly, by substituting all variances into expression \eqref{eq:35}, we obtain the SINR defined in \eqref{eq:17}.

\section{Intersection probability} \label{app:b}
The derivation of the probability $\varphi_{i} = \textrm{Pr}\{\gamma_j = 0, \forall j: o_j \leq o_i \, \textrm{and} \, \gamma_{j: o_j = o_i + 1} = 1 \vert \mathbf{z}\}$ is not trivial, given the statistical dependence between $\gamma_j$. In particular, since these RVs are functions of the SINRs in \eqref{eq:17}, apart from the noise $w$, they will share a part of the interference\footnote{For instance, when considering $M = 3$ and $\mathbf{o} = [1,2,3]$, the received signals coming from sensors $1$ and $2$ will be both affected by the same component from sensor $3$, represented by $\hat{\mathbf{h}}_3 s_3$ (cf. Appendix~\ref{app:a}).}. 
 
For simplicity, in the following, we assume a decoding order $\mathbf{o}=[1,\ldots, M]$. As a result, the probability of correctly retrieving the messages from sensors $j \leq i$ can be written as:
\setcounter{equation}{0}
\begin{equation}
    \varphi_{i} = \textrm{Pr}\{\gamma_1 = 0 \vert \mathbf{z}\} \textrm{Pr}\{\gamma_2 = 0 \vert \gamma_1 = 0, \mathbf{z}\} \times \cdots \times \textrm{Pr}\{\gamma_{i+1} = 1 \vert \gamma_1 = 0, \ldots,\gamma_i = 0, \mathbf{z}\},
\label{eq:41}
\end{equation}
where $\textrm{Pr}\{\gamma_1 = 0 \vert \mathbf{z}\}$ is $1 - \textrm{PER}_1 \vert \mathcal{S}_1, \mathbf{z}$, yet the rest of the conditional probabilities depend on the MCS, the previous decodings\footnote{In fact, these probabilities differ from the corresponding PERs, e.g., $ \textrm{Pr}\{\gamma_2 = 0 \vert \gamma_1 = 0, \mathbf{z}\} \geq  1 - \textrm{PER}_2 \vert \mathcal{S}_2, \mathbf{z}$, since the correct decoding of sensor $1$ ($\gamma_1 = 0$) provides information on the remaining (common) interfering terms $\hat{\mathbf{h}}_j s_j$ for $j \geq 3$.} and an analytic closed-form expression is not usually available. Recall that, in continuous communications ($n_i \to \infty$), the condition $\gamma_j = 0$ translates into the SINR being above a threshold; i.e., the channel has a binary behavior (either you are in an outage or not). However, in the packet-based setup, finding an analogous statement might be difficult (please see the discussion in \cite[Section~V]{Sch20}).

Alternatively, we can consider all previous events to be statistically independent, provided that, in many realistic scenarios, each sensor will use a different coding scheme (rate and structure). Thus, we can safely assume that the CN observes different realizations of this shared interference when decoding each sensor, i.e., the portion of the signal from sensor $i$ that interferes with the sensors $j < i$ will differ for each device. Intuitively, the different codes see this common part from different points of view, and thus, the interference is virtually randomized. The same effect could be achieved with the help of an interleaver \cite{Tse05}.

This way, thanks to the presence of the channel decoder, the set of probabilities $\varphi_{i}$ result in:
\begin{equation}
    \varphi_{i} \approx \left(\textrm{PER}_{i+1} \vert \mathcal{S}_{i+1},\mathbf{z}\right) \prod_{j = 1}^i \left(1 - \textrm{PER}_j \vert \mathcal{S}_j, \mathbf{z} \right),
    \label{eq:42}
\end{equation}
which essentially represents the decoding procedure described in Section~\ref{sec:4}, i.e., sensor $i$ can only be decoded if (and only if) the previous sensors are correctly decoded (but not completely suppressed due to I-CSI errors). Otherwise, we consider that the decoding is not possible and, thus, the information from that sensor cannot be recovered.

\section{Observation statistics} \label{app:c}
Under the assumption of Rician channels (refer to Subsection~\ref{sec:6.1}), the vectorized cascaded channel can be expressed as follows:
\setcounter{equation}{0}
\begin{equation}
    \begin{aligned}
        \mathbf{g}_{\textrm{C},i} &= c_i \textrm{vec}\Big( \bm{\Delta}_{\textrm{R}}\textrm{diag}\left(\bm{\tau}_i\right) + \sqrt{F_1}\bm{\Delta}_{\textrm{R}}\textrm{diag}\left(\mathbf{v}_i\right) \\
        &\quad + \sqrt{F_2} \mathbf{u}_{\textrm{R}}\mathbf{v}_{\textrm{R}}^{\textrm{H}}\textrm{diag}\left(\bm{\tau}_i\right) + \sqrt{F_1F_2}\mathbf{u}_{\textrm{R}}\mathbf{v}_{\textrm{R}}^{\textrm{H}}\textrm{diag}\left(\mathbf{v}_i\right)\Big),    
    \end{aligned}    
    \label{eq:43}
\end{equation}
with $c_i \triangleq \sqrt{\frac{(\delta_i\delta_{\textrm{R}})^{-\alpha_2}}{(1 + F_1)(1 + F_2)}}$, $\mathbf{v}_i \triangleq \mathbf{v}(\vartheta_i)$, $\mathbf{v}_{\textrm{R}} \triangleq \mathbf{v}(\vartheta_{\textrm{R},2})$, and $\mathbf{u}_{\textrm{R}} \triangleq \mathbf{u}(\vartheta_{\textrm{R},1})$. The first term represents the product of two complex and independent Gaussian RVs, the second and third terms follow a complex Gaussian distribution, and the last term is a constant.

When considering the observations for sensor $i$, we obtain 
\begin{equation}
    \begin{aligned} 
        \mathbf{z}_i &= c_i\underbrace{\mathbf{S}\textrm{vec}\left( \bm{\Delta}_{\textrm{R}}\textrm{diag}\left(\bm{\tau}_i\right)\right)}_{\triangleq \mathbf{z}_{i,1}} + c_i\sqrt{F_1} \underbrace{\mathbf{S} \textrm{vec}\left(\bm{\Delta}_{\textrm{R}}\textrm{diag}\left(\mathbf{v}_i\right)\right)}_{\triangleq \mathbf{z}_{i,2}} + c_i\sqrt{F_2}\underbrace{\mathbf{S} \textrm{vec}\left(\mathbf{u}_{\textrm{R}}\mathbf{v}_{\textrm{R}}^{\textrm{H}}\textrm{diag}\left(\bm{\tau}_i\right)\right)}_{\triangleq \mathbf{z}_{i,3}} \\
        &\qquad + c_i\sqrt{F_1F_2}\mathbf{S}\textrm{vec}\left(\mathbf{u}_{\textrm{R}}\mathbf{v}_{\textrm{R}}^{\textrm{H}}\textrm{diag}\left(\mathbf{v}_i\right)\right) + \frac{1}{N P_i} \left(\left(\mathbf{I}_T \otimes \mathbf{p}_i\right)^{\textrm{T}}\otimes \mathbf{I}_K\right)\textrm{vec}\left(\mathbf{W}\right),
    \end{aligned}
    \label{eq:44}
\end{equation}
where $\mathbf{z}_{i,2} \sim \mathcal{CN}(\mathbf{0}_{KT},\mathbf{S}\mathbf{S}^{\textrm{H}})$ and $\mathbf{z}_{i,3} \sim \mathcal{CN}(\mathbf{0}_{KT},\mathbf{S}(\mathbf{I}_L \otimes \mathbf{u}_{\textrm{R}}\mathbf{u}_{\textrm{R}}^{\textrm{H}})\mathbf{S}^{\textrm{H}})$ (refer to \eqref{eq:9}). Note that $\mathbf{W} \triangleq [\mathbf{W}_1,\ldots,\mathbf{W}_T]$ for brevity in the notation.

For clarity in the explanation, let us proceed with $T = K = 1$, i.e., $\mathbf{S} = \bm{\upsilon}^{\textrm{T}}$. Hence, the first term $\mathbf{z}_{i,1}$ is represented by the sum of independent (scalar) RVs:
\begin{equation}
    \mathbf{z}_{i,1} = \sum_{l = 1}^L \upsilon_l \bm{\Delta}_{\textrm{R},l} \bm{\tau}_{i,l},
    \label{eq:45}
\end{equation}
which, in accordance with the central limit theorem, tends to a complex Gaussian RV with zero mean and variance $\sum_l \vert \upsilon_l \vert^2 = L$, as the number of reflecting elements increases \cite{Yil21}. Therefore, it is evident that $\mathbf{z}_{i,1} \sim \mathcal{CN}(\mathbf{0}_{KT},\mathbf{S}\mathbf{S}^{\textrm{H}})$. 

Since the noise is zero mean and independent of the other RVs, $\mathbf{z}_i$ also follows a complex Gaussian distribution with mean $\bm{\mu}_{\mathbf{z}_i} = c_i \sqrt{F_1F_2} \mathbf{S}\textrm{vec}(\mathbf{u}_{\textrm{R}}\mathbf{v}_{\textrm{R}}^{\textrm{H}}\textrm{diag}(\mathbf{v}_i))  \in \mathbb{C}^{KT}$ and covariance matrix
\begin{equation}    
    \mathbf{C}_{\mathbf{z}_i} = c_i^2 \mathbf{S}\left(\left(1 + F_1\right)\mathbf{I}_{KL} + F_2\left(\mathbf{I}_L \otimes \mathbf{u}_{\textrm{R}}\mathbf{u}_{\textrm{R}}^{\textrm{H}}\right))\right)\mathbf{S}^{\textrm{H}} + \frac{\sigma_w^2}{N P_i} \mathbf{I}_{KT}  \in \mathbb{C}^{KT \times KT}.    
    \label{eq:46}
\end{equation}

When considering the entire set of observations, all that remains is to compute the cross-covariance matrices $\mathbf{C}_{\mathbf{z}_i,\mathbf{z}_j} \in \mathbb{C}^{KL \times KL}$. In this case, the only common and random part between the observations from different sensors is $\bm{\Delta}_{\textrm{R}}$. Hence, since $\bm{\Delta}_{\textrm{R}}$ and $\bm{\tau}_i$ are both zero mean and independent (meaning that the cross-products cancel out), it can be shown that:
\begin{equation}
    \mathbf{C}_{\mathbf{z}_i,\mathbf{z}_j} = F_1 c_i c_j  \mathbf{S} \left(\textrm{diag}(\mathbf{v}_i)\textrm{diag}(\mathbf{v}_j)^{\textrm{H}} \otimes \mathbf{I}_K\right)\mathbf{S}^{\textrm{H}}.
    \label{eq:47}
\end{equation}

\section{UatF SINR} \label{app:d}
The SINR derived from the also popular, yet looser, UatF or \textit{hardening} lower bound for the data rate, first introduced in \cite{Med00} relies on the premise that only the mean of the effective channel is known \cite{Bjo17}. Yet, the rest is treated as uncorrelated Gaussian noise. Consequently, the signal model is given by 
\setcounter{equation}{0}
\begin{equation}
    \begin{aligned}
        \bar{s}_i &\triangleq \mathbb{E}\left[\mathbf{f}_i^{\textrm{H}}\mathbf{h}_i\right] s_i + \left(\mathbf{f}_i^{\textrm{H}}\mathbf{h}_i - \mathbb{E}\left[\mathbf{f}_i^{\textrm{H}}\mathbf{h}_i\right]\right)s_i \\        
        \qquad& + \sum_{j : o_j < o_i}\left(\mathbf{f}_i^{\textrm{H}}\mathbf{h}_j -  \mathbb{E}\left[\mathbf{f}_i^{\textrm{H}}\mathbf{h}_j\right]\right)s_j + \sum_{j : o_j > o_i} \mathbf{f}_i^{\textrm{H}} \mathbf{h}_j s_j + \mathbf{f}_i^{\textrm{H}}\mathbf{w},    
    \end{aligned}    
    \label{eq:48}
\end{equation}
and following the steps outlined in Appendix~\ref{app:a}, the corresponding SINR is
\begin{equation}
    \rho_i \vert \mathcal{S}_i = \frac{P_i \left\vert \mathbb{E}\left[\mathbf{f}_i^{\textrm{H}}\mathbf{h}_i\right] \right\vert^2}{\displaystyle \sigma_w^2 \mathbb{E}\left[\left\| \mathbf{f}_i \right\|_2^2\right] + \sum_{j : o_j \leq o_i}P_j\mathbb{E}\left[\left\vert\mathbf{f}_i^{\textrm{H}}\mathbf{h}_j -  \mathbb{E}\left[\mathbf{f}_i^{\textrm{H}}\mathbf{h}_j\right] \right\vert^2\right] + \sum_{j : o_j > o_i} P_j \mathbb{E}\left[\left\vert\mathbf{f}_i^{\textrm{H}} \mathbf{h}_j \right\vert^2\right]}.
    \label{eq:49}
\end{equation}

In contrast to previous cases, conditioning on the set of observations $\mathbf{z}$ is unnecessary since we rely exclusively on CDI, allowing for the derivation of a closed-form expression. Despite the UatF approach yielding poorer performance, as discussed in Subsection~\ref{sec:6.2}, it significantly reduces computational complexity.

More specifically, considering the use of MRC or matched filter processing along with the MMSE channel estimation (cf. Section~\ref{sec:3}), the first-order moment\footnote{To avoid redundancy, only the moments of the desired signal are presented here. Similar derivations hold for the rest of the cross-terms.} is
\begin{equation}
    \mathbb{E}\left[\mathbf{f}_i^{\textrm{H}}\mathbf{h}_i\right] = \mathbb{E}\left[\hat{\mathbf{h}}_i^{\textrm{H}}(\hat{\mathbf{h}}_i + \tilde{\mathbf{h}}_i)\right] = \mathbb{E}\left[\hat{\mathbf{h}}_i^{\textrm{H}}\hat{\mathbf{h}}_i\right] = \textrm{tr}\left(\widehat{\mathbf{C}}_i\right) + \hat{\bm{\mu}}_i^{\textrm{H}}\hat{\bm{\mu}}_i,
    \label{eq:50} 
\end{equation}
where $\hat{\bm{\mu}}_i \triangleq \mathbb{E}[\hat{\mathbf{h}}_i] \in \mathbb{C}^{K}$ and $\widehat{\mathbf{C}}_i \triangleq \mathbb{E}[(\hat{\mathbf{h}}_i - \hat{\bm{\mu}}_i)(\hat{\mathbf{h}}_i - \hat{\bm{\mu}}_i)^{\textrm{H}}] \in \mathbb{C}^{K \times K}$ represent the mean vector and the covariance matrix of the estimated effective channel, respectively. 

Following the reasoning in Appendix~\ref{app:c}, owing to the independence between direct and cascaded channels, we have\footnote{The inverse of the vectorized form $\mathbf{x} \triangleq \textrm{vec}(\mathbf{X})$ of any matrix $\mathbf{X} \in \mathbb{C}^{m \times n}$ can be obtained using the expression $(\textrm{vec}(\mathbf{I}_n)^{\textrm{T}} \otimes \mathbf{I}_m)(\mathbf{I}_n \otimes \mathbf{x})$.}
\begin{align}
    \label{eq:51} \hat{\bm{\mu}}_i &= (\textrm{vec}(\mathbf{I}_L)^{\textrm{T}} \otimes \mathbf{I}_K)(\mathbf{I}_L \otimes \bm{\mu}_{\hat{\mathbf{g}}_{\textrm{C},i}}) \bm{\psi}, \\
    \label{eq:52} \widehat{\mathbf{C}}_i &= d_i^{-\alpha_1}\mathbf{I}_K + \begin{bmatrix} \textrm{tr}\left(\mathbf{C}_{\hat{\mathbf{g}}_{\textrm{C},i}}\mathbf{E}_{1,1}\right) & \ldots & \textrm{tr}\left(\mathbf{C}_{\hat{\mathbf{g}}_{\textrm{C},i}}\mathbf{E}_{K,1}\right) \\ \vdots & \ddots & \vdots \\  \textrm{tr}\left(\mathbf{C}_{\hat{\mathbf{g}}_{\textrm{C},i}}\mathbf{E}_{1,K}\right) & \ldots & \textrm{tr}\left(\mathbf{C}_{\hat{\mathbf{g}}_{\textrm{C},i}}\mathbf{E}_{K,K}\right) \end{bmatrix},
\end{align}
where $\mathbf{E}_{k,k'} \triangleq \bm{\psi}\bm{\psi}^{\textrm{H}}\otimes \mathbf{e}_{k}\mathbf{e}_{k'}^{\textrm{T}}$.

Unfortunately, the second-order moment can become tedious and cumbersome due to the non-circular symmetry of the RVs (i.e., the channels are not zero mean) \cite{Zhi23}. To address this, we apply the inequality $\vert a + b \vert^2 \leq 2(\vert a \vert^2 + \vert b \vert^2)$, yielding the following upper bounds\footnote{It is crucial to emphasize that since these terms are in the denominator, these bounds will represent worst-case scenarios.}:
\begin{equation}
    \begin{aligned}
        \mathbb{E}\left[\left\vert \mathbf{f}_i^{\textrm{H}}\mathbf{h}_i \right\vert^2\right] &=  \mathbb{E}\bigg[\Big\vert(\underbrace{\hat{\mathbf{h}}_i - \hat{\bm{\mu}}_i}_{\triangleq \check{\mathbf{h}}_i} + \hat{\bm{\mu}}_i)^{\textrm{H}}(\underbrace{\mathbf{h}_i - \bm{\mu}_i}_{\triangleq \bar{\mathbf{h}}_i} + \bm{\mu}_i)\Big\vert^2\bigg] \\
        &= \mathbb{E}\left[\left\vert\check{\mathbf{h}}_i^{\textrm{H}}\bar{\mathbf{h}}_i + \check{\mathbf{h}}_i^{\textrm{H}}\bm{\mu}_i + \hat{\bm{\mu}}_i^{\textrm{H}}\bar{\mathbf{h}}_i + \hat{\bm{\mu}}_i^{\textrm{H}}\bm{\mu}_i\right\vert^2\right] \\
        &\leq 2 \, \Big(\mathbb{E}\left[\left\vert\check{\mathbf{h}}_i^{\textrm{H}}\bar{\mathbf{h}}_i \right\vert^2\right] + \left\vert \hat{\bm{\mu}}_i^{\textrm{H}}\bm{\mu}_i\right\vert^2 
        + 2 \, \textrm{Re}\left\{\mathbb{E}\left[\check{\mathbf{h}}_i^{\textrm{H}}\bar{\mathbf{h}}_i\right]\bm{\mu}_i^{\textrm{H}}     \hat{\bm{\mu}}_i\right\} \\
        &\qquad + \bm{\mu}_i^{\textrm{H}}\mathbb{E}\left[\check{\mathbf{h}}_i\check{\mathbf{h}}_i^{\textrm{H}} \right]\bm{\mu}_i + \hat{\bm{\mu}}_i^{\textrm{H}}\mathbb{E}\left[\bar{\mathbf{h}}_i\bar{\mathbf{h}}_i^{\textrm{H}}\right]\hat{\bm{\mu}}_i\Big),
    \end{aligned}
    \label{eq:53}
\end{equation}
where $\bm{\mu}_i \triangleq \mathbb{E}[\mathbf{h}_i] \in \mathbb{C}^{K}$ is the mean of the effective channel:
\begin{equation}
     \bm{\mu}_i = (\textrm{vec}(\mathbf{I}_L)^{\textrm{T}} \otimes \mathbf{I}_K)(\mathbf{I}_L \otimes \bm{\mu}_{\mathbf{g}_{\textrm{C},i}}) \bm{\psi}.
     \label{eq:54} 
\end{equation}

Finally, as the zero-mean channels $\check{\mathbf{h}}_i$ and $\bar{\mathbf{h}}_i$ are jointly Gaussian (please refer to Appendix~\ref{app:c} for more details), it can be shown that \cite{Neu18}
\begin{equation}
    \mathbb{E}\left[\left\vert \check{\mathbf{h}}_i^{\textrm{H}}\bar{\mathbf{h}}_i\right\vert^2\right] = \textrm{tr}\left(\mathbb{E}\left[\check{\mathbf{h}}_i\check{\mathbf{h}}_i^{\textrm{H}}\right]\mathbb{E}\left[\bar{\mathbf{h}}_i \bar{\mathbf{h}}_i^{\textrm{H}}\right]\right) + \left\vert \textrm{tr}\left(\mathbb{E}\left[\bar{\mathbf{h}}_i \check{\mathbf{h}}_i^{\textrm{H}}\right]\right)\right\vert^2 = \textrm{tr}\left(\widehat{\mathbf{C}}_i \mathbf{C}_i\right) + \left\vert \textrm{tr}\left(\widehat{\mathbf{C}}_i\right)\right\vert^2,    
    \label{eq:55}
\end{equation}
where $\mathbf{C}_i \triangleq \mathbb{E}[(\mathbf{h}_i - \bm{\mu}_i)(\mathbf{h}_i - \bm{\mu}_i)^{\textrm{H}}] \in \mathbb{C}^{K \times K}$ is the covariance matrix of the actual value $\mathbf{h}_i$, obtained by replacing $\mathbf{C}_{\hat{\mathbf{g}}_{\textrm{C},i}}$ with $\mathbf{C}_{\mathbf{g}_{\textrm{C},i}}$ in \eqref{eq:52}.

\end{appendices}

\newpage

\section*{Abbreviations}

\begin{table*}[h]        
    \begin{tabular}{l l}            
        \textbf{AO} & Alternating optimization
        \\
        \textbf{AoA} & Angle of arrival
        \\        
        \textbf{AWGN} & Additive white Gaussian noise        
        \\            
        \textbf{BS} & Base station
        \\
        \textbf{CDI} & Channel distribution information
        \\            
        \textbf{CN} & Collector node
        \\
        \textbf{CSI} & Channel state information (perfect)
        \\            
        \textbf{FBL} & Finite blocklength
        \\
        \textbf{I-CSI} & Imperfect CSI
        \\
        \textbf{LMMSE} & Linear MMSE
        \\
        \textbf{LoS} & Line of sight                    
        \\
        \textbf{MCS} & Modulation and coding scheme
        \\
        \textbf{MMSE} & Minimum mean square error
        \\
        \textbf{MRC} & Maximum ratio combining
        \\
        \textbf{MTC} & Machine-type communications                        
        \\
        \textbf{NLoS} & Non-LoS
        \\
        \textbf{NMSE} & Normalized MSE
        \\
        \textbf{NOMA} & Non-orthogonal multiple access            
        \\
        \textbf{OFDM} & Orthogonal frequency-division multiplexing
        \\
        \textbf{PER} & Packet error rate
        \\
        \textbf{RIS} & Reconfigurable intelligent surface
        \\
        \textbf{RV} & Random variable
        \\
        \textbf{SIC} & Successive interference cancelation                    
        \\
        \textbf{SINR} & Signal-to-interference-plus-noise ratio       
        \\
        \textbf{SNR} & Signal-to-noise ratio                
        \\
        \textbf{THz} & Terahertz                                    
        \\
        \textbf{UatF} & Use-and-then-forget
        \\
        \textbf{UL} & Uplink        
    \end{tabular}                        
\end{table*}

\section*{Declarations}

\bmhead{Acknowledgements}
Not applicable.

\bmhead{Authors' contributions}
All authors contributed equally to this work.

\bmhead{Authors' information}
This work was conducted when Sergi Liesegang was with the Signal Processing and Communications Group, recognized as a consolidated research group by the Departament de Recerca i Universitats de la Generalitat de Catalunya through 2021 SGR 01033, at the Department of Signal Theory and Communications from the Universitat Polit\`ecnica de Catalunya.

\bmhead{Availability of data and material} 
The datasets analyzed during the current study are available in the Intel Lab Data repository (\url{https://db.csail.mit.edu/labdata/labdata.html}).

\bmhead{Competing interests}
Any author has financial/non-financial conflict of interest.

\bmhead{Funding}
This work was supported in part by the project ROUTE56 under the grant PID2019-104945GB-I00, funded by MCIN/AEI/10.13039/501100011033; the project 6-SENSES under the grant PID2022-138648OB-I00, funded by MCIN/AEI/10.13039/501100011033 and FEDER-UE, ERDF-EU “A way of making Europe”; and the scholarship FPI under the grant BES-2017-079994, funded by the Spanish Ministry of Science, Innovation, and Universities.

\bibliography{./references}

\end{document}